\documentclass[9.5pt,journal,compsoc]{IEEEtran}
\AtBeginDocument{%
  \providecommand\BibTeX{{%
    \normalfont B\kern-0.5em{\scshape i\kern-0.25em b}\kern-0.8em\TeX}}}

\usepackage{xcolor} 
\usepackage{listings}
\usepackage{pifont}
\usepackage{todonotes}
\usepackage{hyperref} 
\usepackage{enumitem}

\usepackage{pgf}
\usepackage{lipsum}
\usepackage{comment}

\usepackage{booktabs}

\usepackage{graphicx}

\usepackage{multirow}
\usepackage{cite}

\usepackage{amsfonts}

\usepackage[noend]{algpseudocode} 

\usepackage[ruled]{algorithm2e} 

\usepackage{url}

\newcommand\thename{\textsc{Mimosa}}

\begin{document}
%
\title{Reducing Malware Analysis Overhead with Coverings}

%

%
\author{\IEEEauthorblockN{
Mohsen Ahmadi\IEEEauthorrefmark{1}
Kevin Leach\IEEEauthorrefmark{2},
Ryan Dougherty\IEEEauthorrefmark{3},
Stephanie Forrest\IEEEauthorrefmark{1} and
Westley Weimer\IEEEauthorrefmark{2}}\\

\IEEEauthorblockA{\IEEEauthorrefmark{1}Arizona State University, Tempe, Arizona
\IEEEauthorblockA{\IEEEauthorrefmark{2}University of Michigan, Ann Arbor, Michigan\\ 
\IEEEauthorblockA{\IEEEauthorrefmark{3}United States Military Academy, West Point, New York\\
\{pwnslinger, stephanie.forrest\}@asu.edu}
~ \{kjleach, weimerw\}@umich.edu}
ryan.dougherty@westpoint.edu}
}



\IEEEtitleabstractindextext{
\begin{abstract}


There is a growing body of malware samples that evade automated analysis and detection tools.
Malware may measure fingerprints ("artifacts") of the underlying analysis tool or environment, and change their behavior when artifacts are detected.
While analysis tools can mitigate artifacts to reduce exposure, such concealment is expensive.
However, not every sample checks for every type of artifact---analysis efficiency can be improved by mitigating only those artifacts most likely to be used by a sample.
Using that insight, we propose \thename{}, a system which identifies a small set of "covering" tool configurations that collectively defeat most malware samples with increased efficiency.
\thename{} identifies a set of tool configurations which maximize analysis throughput and detection accuracy while minimizing manual effort, enabling scalable automation for analyzing stealthy malware.

We evaluate our approach against a benchmark of 1535 labeled stealthy malware samples. 
Our approach increases analysis throughput over the state of the art on over 95\% of these samples.
We also investigate cost-benefit tradeoffs between the fraction of successfully-analyzed samples and computing resources required.
\thename{} provides a practical, tunable method for efficiently deploying analysis resources.

\end{abstract}

}






\maketitle

%


\section{Introduction}

Malware continues to proliferate, significantly eroding user and
corporate privacy and trust in computer
systems~\cite{kaspersky-report-2017,mcafee1q-2016report,mcafee4q-2014report,symantec-istr2017}.
Malwarebytes Threat Landscape reported a 13\% increase in malware targeting
businesses in 2019~\cite{malwarebytes-report-2020}. SonicWall detected around
10 billion malware attacks in 2019~\cite{sonicwall-report-2020}.  Although Symantec notes
a 61\%  decrease in the number of new malware variants
between 2017 and 2018, the distribution of specific samples like Adware/InstallCore increased 360\% from
2018 to 2019~\cite{symantec-report-2019, malwarebytes-report-2020}.  
Keeping abreast of this large 
volume of malware requires effective scalable malware analysis techniques. 


Once a malware sample has been detected and analyzed, automated techniques such as signature matching can quickly identify other copies. Understanding novel malware samples, however, requires lengthly 
analysis using both automated and manual techniques~\cite{malware-analysis1, forensic-discovery}.  
Analysts frequently execute samples under laboratory
setups~\cite{sans-report, langeland2008} using virtualization.  
This includes not only virtual machine monitors like VMWare~\cite{vmware},
Xen~\cite{xen}, and VirtualBox~\cite{virtualbox}, but also tools that depend on
virtualization such as Ether~\cite{ether}, HyperDbg~\cite{hyperdbg:ase10}, or
Spider~\cite{spider:acsac13}.  Executing the malware sample in a controlled environment allows the analyst to observe its behavior safely.  If malware causes damage,the damage is limited to
the virtualized environment, which can be destroyed and restarted to
analyze subsequent samples.  Virtualization is now a lynchpin of computer security and analysis
applications~\cite{draugr, volatility, hay2008forensics, jiang2007stealthy,
fatkit}.

As these malware analysis methods have matured, malware authors have in turn adopted evasive, or \textit{stealthy}, techniques to avoid or subvert automated analysis~\cite{sec-week-18,anti-debug-chen-DSN08,lophi}.  
Chen \emph{et al.}~\cite{anti-debug-chen-DSN08}, for example,
reported that 40\% of malware samples hide or reduce
malicious behavior when run in a VM or with a debugger attached. Stealthy malware
techniques include anti-debugging~\cite{anti-debug-chen-DSN08, anti-debug-bh12,
anti-debug-symantec}, anti-virtualization~\cite{anti-vm-ldt,anti-vm-exception}, and
anti-emulation~\cite{anti-emulator-ISC07}.  These methods
detect a particular feature, or \emph{artifact}, of
the analysis environment which allows the malware to determine if it is being
analyzed. When an artifact is detected, the malware can avoid
executing its malicious payload, thereby hiding its true function from the
analyst. Table~\ref{tab:artifacts} summarizes common artifacts, derived from Zhang \emph{et
al.}~\cite{malt-sp15}.  Studying the behavior of stealthy malware requires that the analyst \textit{mitigate} the artifacts by configuring the environment in a way that prevents detection by the malware.
Over time, malware authors have discovered a wide diversity of artifact types, which has increased the time required to
manually determine the best mitigation strategy for each malware sample.  This process 
has proven difficult to
automate.  


Given current trends, we expect that malware authors will continue discovering new artifacts, forcing analysts to develop new
mitigations, leading to continued escalation of the complexity and cost of
conducting malware analysis.
Balzarotti \emph{et al.} describe how stealthy
malware samples check for evidence of analyses and behave differently when
they are present~\cite{balzarotti2010efficient}. They classify stealthy
malware by running samples under multiple environments and using the
differences between those runs, especially in terms of patterns of system
call execution, to characterize evasive behavior. For example, if a sample
is executed under both VMWare~\cite{vmware} and VirtualBox~\cite{virtualbox}, and the VMWare instance does
not exhibit malicious behavior, one can conclude that the sample detects
VMWare-specific artifacts (e.g.,~\cite{oyama2018trends}). Many
techniques, from machine learning~\cite{pan2017dark} to symbolic
execution and traces~\cite{hong2018} to hybrid dynamic
analyses~\cite{qiang2018}, among others, have been proposed
to tackle this problem of environment-aware malware---even as
new black hat approaches for more insidious stealthy evasion
(e.g.,~\cite{tanabe2018,miramir2017}) are proposed as well. 

This paper presents \thename\footnote{\thename{}: Malware Instrumentation with Minimized Overhead for Stealthy Analysis.} 
to address the need for high-throughput, low-overhead
automated analysis of stealthy malware.  \thename{}'s key insight is that any malware sample is likely to use a small set of artifact mitigation strategies out of the large set of possible mitigations.  
We propose using
\emph{coverings} to find a small set of analysis
configurations that collectively cover (mitigate) the techniques used by most stealthy malware samples
while minimizing the cost of each individual analysis configuration.
\thename{} can be used as part of an automated malware analysis or triage system to help detect and understand new stealthy malicious samples.

We extend the previous state-of-the-art to consider both
the \emph{cost} and \emph{coverage} of artifact mitigation strategies. 
Given the popularity of stealthy malware and the increasing number of
anti-stealth techniques, the question is no longer whether or not evasion
\emph{should} be mitigated, but \emph{which set} of techniques should be used for a particular sample.
Since samples often use combinations of artifacts to evade detection~\cite{sec-week-18}, this
is not a simple decision. 
First, each stealth mitigation technique comes with
associated costs---development time, deployment time, CPU time, memory and disk utilization, runtime overhead, etc.---compared
to a bare-metal or bare-VM setup. 
These costs are critical because the rate at which
new malware is deployed~\cite{tanabe2018evasive} combined with the
time and resources required to complete each analysis has led to a situation in which 
analysis time can be a
bottleneck~\cite{bulazel2017survey}. Second, some stealth mitigation techniques
supplant or subsume others but with different costs. For example,
an API call can be hooked to read VMWare-specific registry keys to prevent malware targeting
that registry key from detecting the environment. Such a strategy is more efficient than using an alternate
approach to hide the registry key.  


\begin{table*} 
    \center \caption{Example artifacts used by stealthy malware~\cite{malt-sp15}.\label{tab:artifacts}}
    \resizebox{\textwidth}{!}{
  \begin{tabular}{ll} 
    \toprule
    \textbf{Artifact Name} & \textbf{Artifact Description} \\ \midrule 
    Hardware ID & VMs have devices with obvious strings (e.g., ``VMWare Hard Drive'') or
    specific identifiers (e.g., MAC address). \\ 
    Registry Key & Windows VMs have telling registry keys
    (e.g., unique dates and times associated with VM creation). \\ 
    CPU behavior & VMs may not faithfully reproduce CPU
    instructions.\\ 
    Resource constraint & Malware analysis VMs may be
    given sparse resources (e.g., $<$20GB hard disk) \\ 
    Timing & VMs may not virtualize internal timers, or may incur noticeable
    overhead\\ 
    Debugger presence & Tools like \texttt{gdb} that
    instrument samples are detectable\\ 
    API calls & API
    calls that are hooked for analysis can be
    detected.\\
    Process names  & VMs, analysis and monitoring tools have some processes with specific predefined names (e.g., vmtoolsd in VMWare). \\ 
    HCI & check the human interactions with system e.g. mouse and keyboard activity \\
  \bottomrule \end{tabular}}

\end{table*}

\noindent To summarize, the main contributions of this paper are: 

\begin{itemize}[leftmargin=5mm]
    \item A new algorithm for identifying a low-cost set of artifact covering configurations;
    \item \thename{}, a system for selecting and deploying covering combinations of artifact mitigations to maximize analysis throughput and accuracy;
    \item An empirical study of 1535 labeled stealthy malware samples from the wild, demonstrating that \thename{} achieves high coverage of stealthy malware and high automated analysis throughput; and
    \item Open-source software that provides a unified framework for conducting scalable evasive malware analysis. We release the codebase of \thename{} under the following Github repository for public access: \href{https://github.com/AdaptiveComputationLab/MIMOSA}{github.com/AdaptiveComputationLab/MIMOSA}.
\end{itemize}

\section{Background}

We call malware \emph{stealthy} if it actively seeks to detect, disable, or
otherwise subvert malware analysis tools.  Stealthy malware operates by checking
for signatures, or \emph{artifacts}, associated with various analysis tools
or techniques.  For example, a malware sample may invoke the
\texttt{isDebuggerPresent} Win32 API call to determine whether a debugger is
attached to the process---if a debugger is attached, the sample may conclude
that an analyst is instrumenting it and change its behavior
accordingly.  There are many different types of artifacts exposed by the wide
variety of analysis tools and frameworks used today.  Briefly, stealthy malware samples
use artifacts as heuristics to determine if they are under analysis, and
change their behavior to subvert the tool.

Stealthy, evasive malware has been studied extensively~\cite{bulazel2017survey}, and is of increasing concern in
industrial settings, with companies such as Minerva and Lastline marketing solutions for
detecting stealthy malware.  
In addition, stealth is often a property gained through the use of packers~\cite{upx,aspack,rlpack} that can systematically change malware statically to evade detection and subvert analysis. 
Thus, there is a need for defensive methods that can keep up with the escalating arms race with malware.

An \emph{artifact} is information about the execution environment that 
a malware sample can use to determine if it is running non-natively.
For example, if a malware sample checks whether a debugger is attached to it, that sample may behave differently in an attempt to conceal its true behavior from an analyst using the debugger.
For any given artifact, there can be multiple
\emph{artifact mitigation strategies} for preventing
exposure of the artifact to the sample.
Each such strategy comes with an associated (1) \emph{mitigation
  cost}, which captures overhead, development effort, or other
economic disadvantage, and (2) generality, or \emph{artifact coverage}, which is
the fraction of stealthy samples defeated by the
strategy.  

We consider three broad malware analysis methods: 

\begin{enumerate}

\item \emph{manual analysis}, in which a human analyst reverse engineers, modifies, and analyzes the sample.  This laborious process can take many hours of effort per sample.

\item \emph{bare metal analysis}, in which the sample is run natively rather than in a VM and thus exposes no artifacts to the sample but also incurs risk to the host environment.

\item \emph{combined environment analysis}, in which the sample is run in multiple disparate environments so that the sample is exposed to disjoint sets of artifacts.

\end{enumerate}

In this paper, we focus on the third approach, namely combined environment analysis.
Earlier work~\cite{balzarotti2010efficient, kirat2014barecloud}
used observed differences between runs in disparate environments to determine
which artifacts are used by a stealthy malware sample.
Historically, however, such approaches have not involved many analysis
environments, instead focusing on case studies that compare runs between
limited numbers of virtualization environments.
Given the growing number of malware mitigation techniques, there is a need for techniques that enable 
fine-grained control over the artifacts exposed by the analysis environment.
By precomputing a set of configurations that can be tested in parallel and reused for different malware samples, we hypothesize that \thename{} will both
increase coverage and analysis scalability of stealthy malware.

\section{Motivating Examples}
\label{sec-motivation} 

In this section, we consider two artifact families commonly used by stealthy
malware to detect an analysis environment: debugger-related API calls and hard
disk capacity. For each artifact family, we highlight multiple mitigation
strategies an analyst might use defeat such evasion, and we illustrate
how each strategy can have a different cost and  effectiveness.  These tradeoffs motivate \thename{}'s design.


At one extreme, the analyst could run the sample on
a bare metal machine without virtualization, exposing the fewest
artifacts (high coverage).
This strategy has high cost because it
precludes parallel analyses involving multi-tenant VMs and it can be
expensive to recover from the malware payload. 
At the other extreme, the analyst might use a single
mitigation strategy (low overhead). 
Recall that stealthy malware operates by executing myriad checks for
such artifacts, sometimes six or more~\cite{sec-week-18}, so this approach 
is likely to have low coverage. Even if the single mitigation strategy
chosen defeats one check, it is unlikely to defeat all of them.
As a third alternative, the analyst could apply every known
mitigation strategy simultaneously. However, in practice, a single sample rarely checks for the majority of known
artifacts.\footnote{Advanced Persistent Threats are an exception, which we exclude from consideration.}
The third alternative is also not practical  because some mitigations are incompatible:
they may require specific VMs or incompatible hardware 
configurations, and combining all possible mitigations will often incur unacceptable overheads.
Given a set of available analysis tools, \thename{} can produce sets of configurations that occupy different points in the cost-coverage space.

\subsection{Debugger Presence Artifact Family}
\label{sec-debugger-presence}

Some stealthy malware samples explicitly check for the presence of 
standard debugging software. 
Analyzing stealthy malware requires tools
that do not expose related artifacts to the malware sample under test.  In some cases,
this can be trivial.  For example, we can mitigate the
\texttt{isDebuggerPresent} API call by hooking it and returning a spoofed value
so that, from the malware's perspective, it appears as though no debugger is
attached.  Such a hook is fairly simple and requires low runtime overhead
(indeed, some debuggers used for malware analysis, such as
OllyDbg~\cite{ollydbg}, offer an option to hook this API call). However,
sometimes this is ineffective: other techniques exist, beyond that single API call,
that can be used by malware to determine the presence of a debugger (e.g.,
\texttt{isRemoteDebuggerPresent} or fields in process control structures).  We
could employ one of many strategies to mitigate this ``debugger presence''
artifact family: 

\begin{enumerate}
    \item do nothing, risking exposure to samples that invoke any API calls related to debugger presence;
    \item hook one or more API calls within the OS;
    \item run in an instrumented virtual machine that does not directly attach a debugger to the sample; or
    \item use a physical machine to preclude exposure.
\end{enumerate}

Strategy (2) is attractive because hooking these API calls would incur
relatively low runtime overhead.  However, hooking API calls like this requires
development effort specific to the platform being used for analysis. Moreover,
deciding to hook API calls may introduce subsequent mechanisms for determining
the presence of a different artifact. For example, hooking API calls in Windows requires modifying
a process data structure, which could itself be checked by the malware.
Alternatively, we could opt to run the sample in another environment such as an
instrumented virtual machine (e.g., Ether~\cite{ether}), but this would incur
more significant runtime overhead, reducing efficiency.  In brief, moving from
strategy (1) to strategy (4) increases the coverage of stealthy malware
samples, but at a greater cost. 

\subsection{Hard Disk Capacity Artifact Family} 
\label{sec-hard-disk-capacity} 
As a more complex example, many stealthy samples will check the size of the
hard disk.  If the hard disk capacity is below some threshold, the sample may
conclude that it is executing in a resource-constrained virtual machine for
automated analysis.  Depending on the guest OS, there may be a variety of API
calls that would, either directly or indirectly, measure the available hard
disk space. Based on our experience, pafish and some other loaders check for a threshold of 60GB, and if the hard drive size is less than this value, they consider it as a potential analysis environment. An analyzer has several strategies for addressing
this ``hard disk capacity'' artifact family:

\begin{enumerate}
 \item do nothing, risking exposure to samples that
look for specific hard drive capacities;
\item hook one or more API calls
associated with disk space;
\item externally hook the API call from a hypervisor context,
\item allocate a larger virtual hard
disk to a virtual machine used for malware analysis; or
\item run the sample on a physical machine to preclude artifact exposure.
\end{enumerate}

Strategy (2) is cheaper in terms of analysis cost, but requires more effort to
research and understand each of the (potentially many) associated API calls
(e.g., in addition to measuring disk size directly by querying disk
information, a malware sample could write a large amount of data and check if
the OS raises an exception once space is depleted). On the other hand,
strategy (4) requires resources and effort at runtime, 
restricting the number of parallel VMs that could be used for malware analysis.
Finally, we could instead allocate an entire physical analysis machine for the
sample, which would successfully mitigate all artifacts in the disk space family for the
largest subset of malware, but also inhibits analysis scalability.   

These examples show how multiple mitigation strategies can exist for the
same artifact family, how those strategies can have different costs, and how those
strategies can vary in coverage or effectiveness. However, although we have 
thus far presented them in linear lists, the conflicts we demonstrated 
mean that a more nuanced representation is merited. For example, for the debugging
presence artifact family, strategies (3) and (4) conflict and cannot be employed
simultaneously. These observations motivate our adoption of the lattice
data structure (to address coverage and conflict concerns) and our
extension of the covering array algorithm (to address coverage and cost
concerns).

\section{Proposed Workflow}
\label{sec-workflow}

\begin{figure}
\center
\includegraphics[width=\columnwidth]{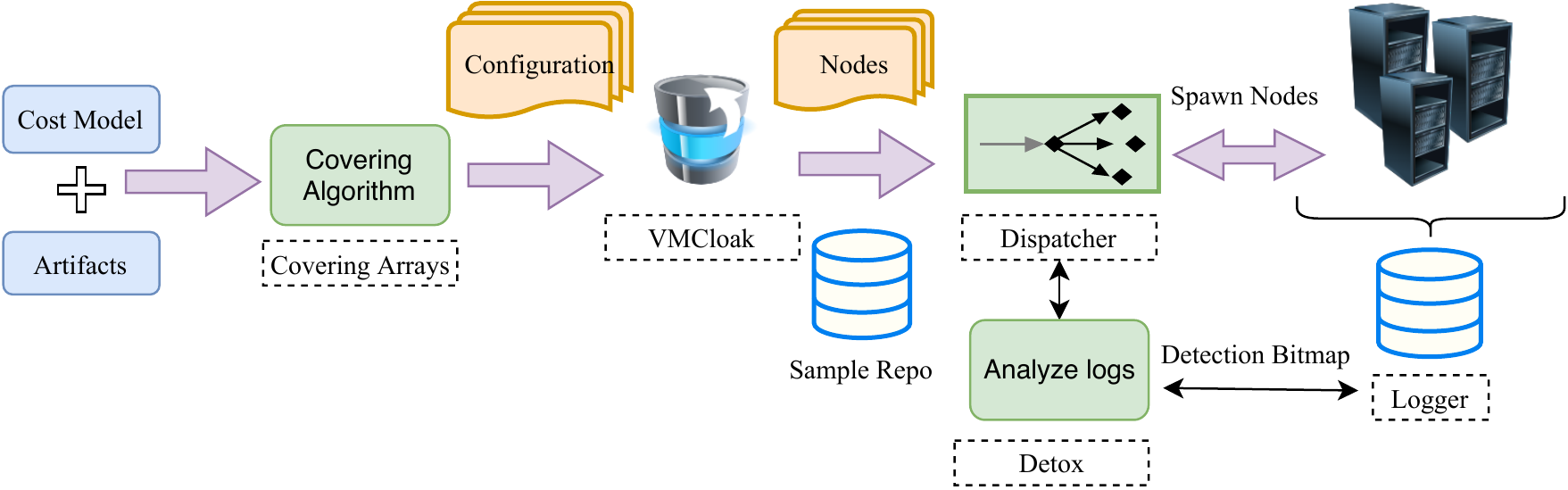}

\caption{A simplified illustration of our \thename{} workflow, consists of four major engines
including Covering algorithm, VMCloak\cite{JBremer2019}, Dispatcher, and Detox. 
\label{fig:simplified_workflow}}

\end{figure}

In this section, we describe the workflow we envision to support.  We seek to
make the automated analysis of stealthy malware more efficient.   Current
techniques either rely on human creativity (e.g., debugging with IDA
Pro~\cite{idapro} or OllyDbg~\cite{ollydbg}) or heavy-weight analysis
techniques that incur significant overhead (e.g., MalT~\cite{malt-sp15} or
Ether~\cite{ether}).  Moreover, differencing approaches, such that of as
Balzarotti \emph{et al.}~\cite{balzarotti2010efficient}, execute a
sample in multiple instrumented environments and use the difference in runs to
determine which artifact is used by the sample, potentially wasting resources.


Given a list of available artifacts, the strategies available for mitigating them, and a cost model, \thename{}'s objective is to select a small set of configurations, which can be deployed in parallel on a given malware sample.  That is, given a fixed number
of available servers, each will be configured to mitigate a different specific subset of artifacts, with lower total cost, e.g., runtime, to lower analysis latency 
compared to existing methods.  Once the covering configurations are identified, \thename{} deploys each of the configurations as a separate instance of an instrumented VM. 
\thename{} manages the VMs to gather logging information and support malware analysis. 

\begin{figure*}
\centering
\includegraphics[width=.95\textwidth]{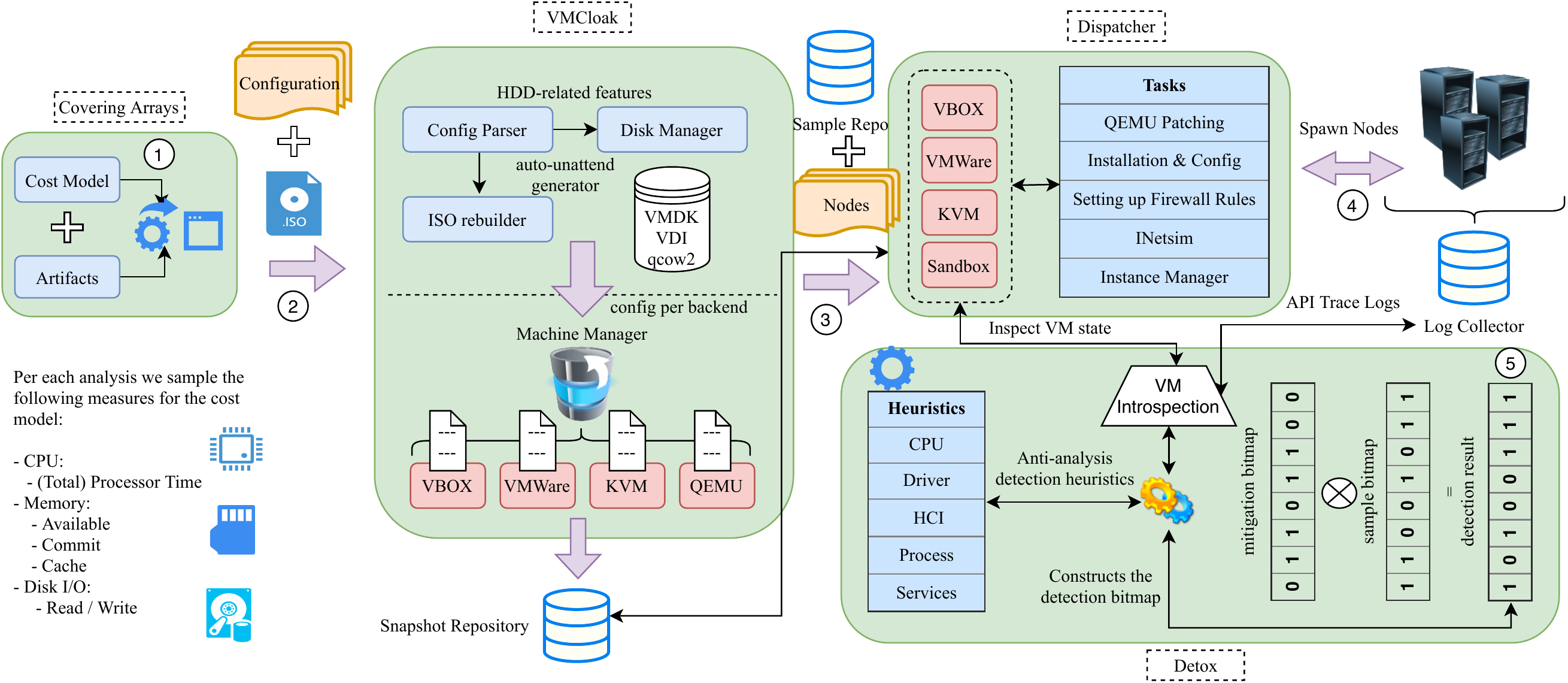}

\caption{
\thename{} workflow: In step~\textcircled{1}, we
develop a set of known artifacts, a set of known mitigation strategies, and a
cost model for each artifact, all of which serve as input to our algorithm.
In step~\textcircled{2}, our covering algorithm generates a set of
mitigation configurations for each server in a particular cluster.  Generated configurations are inputs to our VMCloak engine that provisions VMs that mitigate subsets of artifacts. 
In step~\textcircled{3}, a malware sample repository, a list of configurations and corresponding VMs are passed to the Dispatcher. 
In step~\textcircled{4}, the Dispatcher spawns and manages the analysis of VM instances based on those provisioned by VMCloak.
We record API call traces, which are analyzed to inspect and monitor VM state. 
In Step \textcircled{5}, our Detox engine correlates the collected API logs and VMI using heuristics to determine whether the malware sample was detected by one of the VM instances or not. 
\label{fig:workflow}}

\end{figure*}

\thename{}'s high-level workflow is illustrated in Figure~\ref{fig:simplified_workflow} with details given in Figure~\ref{fig:workflow}.  In Step 1, we apply our covering algorithm (Algorithm~\ref{alg:pareto_frontier}),
which takes (1) a list of artifacts,  (2) corresponding
costs for each artifact (Section~\ref{sec-cost-model}), and (3) a set of mitigation strategies for each artifact (Section~\ref{sec-make-lattice}) as input.
The algorithm returns a covering set of configurations for designing and deploying different virtual machines.  Each covering is represented as a vector of bits, where each element indicates whether that artifact should
be mitigated in the server's configuration. 


The cost model can include a multitude of factors, as determined by the analyst, including VM run-time, memory usage, development time of the mitigation, etc.
\thename{}
uses our covering algorithm (Section~\ref{sec:generating_coverings}) to determine a
set of mitigation configurations for each server in a malware analysis cluster based on the cost model.

Next, these coverings are realized in a malware analysis cluster by configuring specific virtual machines. 
The VMCloak module receives the set of configurations generated previously and
maps VM snapshots to nodes in the cluster.
VMCloak is \thename{}'s custom VM provisioning and deployment framework.
Coverings may entail specific virtualization backends (e.g., QEMU vs. VirtualBox), hooking API calls, 
or modifying the guest kernel (e.g., network drivers). 

With a configuration established for each server in the analysis cluster, \thename{} next
allocates samples to servers.  We assume access to a suite
of hypervisors and hardware resources that can be configured a priori to realize the set of
mitigations specified by the covering.  As described in 
Section~\ref{sec:artifact_mitigation}, we
implemented 13 such hypervisor and hardware configurations
(Table~\ref{tab-config}), managed by our dispatcher module to spin up and spin down analysis
resources as samples are processed.

Each sample is then executed as a process within each configuration.  As the
process runs, \thename{} collects API traces and VM state logs through Virtual Machine Introspection (VMI) and determines
heuristically determine whether the sample is executed successfully.  These heuristics are stored in the Detox
engine, which correlates process and VM
execution logs to infer more semantic patterns. 
In particular, Detox detects if the process
exits, if certain network communication patterns exist, or if certain process
names are created.  We conclude that a sample has been executed successfully if it runs
to termination under each environment.  
Section~\ref{sec-evaluation} uses well-labeled corpus of stealthy
malware to evaluate \thename{}'s effectiveness.

\section{Algorithmic Approach} 
\label{sec-approach}

A key insight of our approach is that efficient 
analysis of malware samples must balance two
competing factors: the number of artifacts that are mitigated and the cost of deploying multiple mitigations.
Because stealthy malware uses artifacts to evade
detection, it is desirable to mitigate as many artifacts as possible to minimize the chance of
disclosing to the sample that it is being analyzed. However, mitigating all artifacts simultaneously imposes unreasonable costs, so the goal is to find sets of \textit{configurations}, where each configuration is a subset of the available artifact mitigations.  The idea then is that each configuration can be run simultaneously, will individually be relatively inexpensive to deploy, but collectively most malware samples will be defeated by at least one configuration.

Given a set of artifact mitigation strategies (configurations) and a model that assigns a cost to each strategy, we describe an algorithm for efficiently selecting a set that maximizes
coverage while minimizing cost. At a high level, there are three main components:

\begin{enumerate}

\item The analyst decides on a {\bf cost model}. Any non-negative cost
function can be used. For example, the model might include development effort and analysis efficiency, combined linearly to compute a total cost.

\item For each artifact family, each mitigation strategy is represented
as a configuration.
Each configuration has an associated cost, computed via the cost model.

\item The {\bf covering algorithm} then selects from the many possible configurations to produce a small set that optimizes the trade-off between cost and coverage. 


\end{enumerate} 

We next describe each component in more detail.

\subsection{Cost Model} 
\label{sec-cost-model}

Abstractly, we model cost as a function mapping each artifact mitigation
strategy to $\mathbb{R}_{\ge0}$. Our approach operates regardless of how this
cost function is defined, but we consider, and provide qualitative details for, two exemplar cost functions: development
time and analysis efficiency (Section~\ref{sec-evaluation}).

If a mitigation strategy is known (e.g., from a published paper) but an
implementation is not available, the analysis organization incurs a software development cost to implement it. 
Software engineers must be paid to
design, implement, test and deploy the mitigation. A full discussion of
software engineering costs is beyond our scope~\cite{huijgens2017effort}, but we note that there are many
organizations or situations in which developer time is an expensive,
limiting resource compared to abundant server, cloud, or compute time. 

Given an available set of implemented mitigations, a second cost is the overhead of deploying them. 
There are a number of relevant metrics here such as
throughput and energy consumption.
Given the rate at which new
stealthy samples are discovered~\cite{tanabe2018evasive,
kirat2014barecloud} and the costs associated with zero-day exploits, rapid
analysis response is often paramount. Given a fixed computing budget, 
if one approach admits analysis after 100 time units and another approach 
only admits analysis after 800 time units, the former is preferred. For
example, consider a scenario in which ten servers are available. One
could deploy heavyweight tools (such as Ether~\cite{ether},
BareCloud~\cite{kirat2014barecloud}, or MalT~\cite{malt-sp15}) on all ten
servers; this would produce a suitable analysis but is not efficient:
it would take a long time for an analysis to run to completion. 
Alternatively, one could deploy lighter-weight systems such as
LO-PHI~\cite{lophi} or VMI-based introspection. This would be more
efficient, but risks detection by samples in the input corpus, at which
point the analysis fails.

Note that while more expensive mitigations usually have higher coverage,
this is not always the case.
For example, if a cost model is used
that captures only developer-hours, then the mitigation strategy of hooking
API calls is both more expensive (it requires a developer to write code)
and less effective than using an alternative analysis environment (which may require little
developer time in such a model).

\subsection{Selecting Artifacts and Mitigations}

\label{sec-make-lattice}

First, we enumerated a number of potential artifacts commonly used by our
corpus of stealthy malware samples on Windows systems (Section~\ref{sec-malware-selection}).  We followed
existing literature~\cite{malt-sp15,bulazel2017survey} and the pafish
tool~\cite{pafish} to group these artifacts into a taxonomy of categories.  We
consider nine artifact families for a total of 39 specific artifacts, which together for a representative 
sample of indicative behavior of stealthy malware.

For each artifact, we implemented several mitigation strategies across a
number of hypervisor backends.  The mitigation strategies ranged in complexity
from straightforward scripting (e.g., synthetic mouse movements) to more
complex patches to the hypervisor source code (e.g., to hook kernel API calls
made within the guest). Table~\ref{tab-awesome} lists each artifacts we considered in our
prototype, and the Appendix shows several example mitigation implementations.  

As new artifacts are discovered in the future and 
exploited by adversaries, mitigations can be implemented and added to \thename{} incrementally.  However, our current implementation includes the artifacts 
exploited by our representative dataset of 1536 manually analyzed stealthy samples. 
The cost analysis and coverings construction, however, generalizes regardless of artifact behavior or exploitation.

\begin{table}
\centering

\caption{Summary of mitigated artifacts in \thename{}.  We categorize artifacts according to conceptual similarity. \label{tab-awesome}}

    \begin{tabular}{p{.48\columnwidth}p{.48\columnwidth}}
\toprule
\textbf{Artifact Family}               & \textbf{Mitigation Examples}                                                                          \\\midrule
 VM-specific Registry Keys             & Hook RegOpenKeyEx API                                                                                     \\
                                       & Hook RegQueryValueEx API                                                                          \\
                                       & Remove offending keys from guest    (e.g., \texttt{HARDWARE\textbackslash ACPI\textbackslash DSDT\textbackslash VBOX\_\_})                                                              \\
                                       & Use alternate VM                                                                                   \\
                                       & Run on bare metal                                                                                  \\


 %
 \midrule
 Mouse / Keyboard / Video Detection        & Spoof peripheral input                                                                                    \\
                                       & Replace spoofed driver files (e.g., VBoxMouse.sys)                                                \\
                                       & Use higher resolution (e.g., $>$800x600)                                                          \\
                                       & Pass through graphics adapter                                                                     \\

 \midrule
 Internal Timing                       & Hook instructions that read MSRs                                                                          \\
                                       & Hook GetLastInputInfo API                                                                         \\
                                       & Hook GetTickCount API                                                                             \\
                                       & Virtualize time stamp counter (TSC)                                                               \\
                                       & Run on bare metal                                                                                 \\

 \midrule
 Device Properties                     & Spoof device names                                                                                        \\
                                       & Allocate bridged network                                                                          \\
                                       & Hook Device Query APIs                                                                            \\
                                       & Hook I/O APIs                                                                                     \\
                                       & Allocate more virtual CPUs                                                                        \\
                                       & Modify BIOS, system, baseBoard, chassis, and OEM Strings                                          \\
                                       & Change NIC MAC address                                                                            \\
                                       & Use alternate VM                                                                                  \\
                                       & Run on bare metal                                                                                 \\

 \midrule
 Drive capacity check                  & Hook CreateFile API                                                                                       \\
                                       & Hook DeviceIoControl API                                                                           \\
                                       & Hook GetDiskFreeSpaceExA  API                                                                      \\
                                       & Hook WriteFile API                                                                                     \\
                                       & Hook GetDriveTypeA API                                                                             \\
                                       & Hook GetVolumeInformationA API                                                                     \\
                                       & Allocate Large virtual disk                                                                        \\
                                       & Allocate physical disk                                                                             \\

 \midrule
 Memory capacity check                 & Hook GlobalMemoryStatusEx API                                                                             \\
                                       & Allocate larger VM guest                                                                              \\
                                       & Run on bare metal                                                                                     \\

 \midrule
 Hooked API detection                  & Externally Hook APIs (e.g., hook hypercalls)                                                                                     \\
                                       & Use hardware breakpoints                                                                          \\
                                       & Run on bare metal                                                                                 \\
 \midrule
 Retrieving CPU Vendor Name            & Patch VMM                                                                                                 \\
                                       & Change VM config                                                                                      \\
                                       & Run on QEMU full system emulation                                                                              \\
 \midrule
 Process / Drive Name Detection        & Patch VMM                                                                                                 \\
                                       & Change VM config                                                                                     \\
                                       & Run on QEMU full system emulation                                                                            \\
 \midrule
 Invalid Instruction Behavior          & Patch VMM                                                                                                 \\
                                       & Use alternate VM                                                                                      \\
                                       & Run on bare metal                                                                                     \\
\bottomrule
\end{tabular}
\end{table}

\subsection{Generating Coverings}
\label{sec:generating_coverings}

Next, we present our algorithm for generating a set of low-cost covers.
Let $\mathcal{A} = \{A_1, \cdots, A_n\}$ be the set of $n$ artifacts, and $\mathcal{C} = \{C_1, \cdots, C_s\}$ be the set of configurations.
Let $S_1, \cdots, S_p$ be the set of samples observed.
For each sample $S_i$, we associate with it a set of \textit{behaviors} $B(S_i)$, which is a subset of the artifacts $\mathcal{A}$.
For each configuration $C_j$, we associate it with a set of \textit{mitigations} $M(C_j)$, which is also a subset of the artifacts $\mathcal{A}$.

Our goal is to construct a binary array (called a \textit{covering}), where each of the rows corresponds to a configuration, and each of the columns corresponds to an artifact, with the following property.
For any sample $S_i$, there are configurations $C_{j_1}, \cdots, C_{j_\ell}$ for which $B(S_i)$ is a subset of $M(C_{j_1}) \cup \cdots \cup M(C_{j_{\ell}})$; in other words, for any sample, there are some configurations that together fully mitigate the sample.
In terms of the array itself, suppose that $B(S_i)$ involves the columns $b_1, \cdots, b_m$. 
Then the union of all rows $r$ in these columns has a 1 in each entry, where 1 in column $b_i$ indicates that configuration $r$ mitigates the artifact $b_i$, and 0 otherwise.
If the property is not maintained, we generate an array that mitigates as many samples as possible (high coverage), while also having the cost(s) of the chosen configurations be as low as possible.

In addition, we maintain a set of  \textit{desirably high} (DH) and \textit{desirably low} (DL) characteristics, where each configuration has a valuation for each of these.
For a covering corresponding to a set of configurations, the measure for the covering of the same characteristic may be the average from each configuration, the total, or some other measure.
For example, if the characteristic is measuring the deployment time, then the total deployment time for a set of configurations is the total over all of their deployment times.
In general, we want to generate a set of configurations such that the covering's DH characteristics are as large as possible, and the DL characteristics are as low as possible.
For the deployment time example, this would be a DL characteristic; coverage would be a DH characteristic.
Because different characteristics can have different impacts on a system, we aim to produce a collection of coverings such that none of them ``overshadow'' any other one.

Next, we walk through the algorithm. 
First, we will discuss generating a covering of all the considered configurations and artifacts relevant for any sample.
For each configuration and artifact, we mark whether or not the configuration mitigates the artifact; the array corresponding to the covering is the natural one.
We determine the cost and coverage of each configuration in turn.
Next, we generate all subsets of configurations; suppose these subsets are $S_1, \cdots, S_k$.
We say that a subset of configurations $S_i$ \textit{dominates} another subset $S_j$ if the following properties hold:
\begin{itemize}
	\item $S_i$'s DH characteristics are all at least those of $S_j$, 
	\item $S_i$'s DL characteristics are all at most those of $S_j$, and
	\item either (1) some DH characteristic of $S_i$ is strictly larger than that of $S_j$, or (2) some DL characteristic of $S_j$ is strictly smaller than that of $S_j$.
\end{itemize}

The \textit{Pareto front} of the subsets is the collection of subsets $\mathcal{S}$ such that none of the subsets dominates any other in $\mathcal{S}$, which can be found by \textit{non-dominated sorting} \cite{deb2002fast}.

This algorithm is not efficient because it examines every subset, which takes exponential time in the number of configurations.
We give an optimization that improves the running time in practice, contingent on the following assumption.
Suppose that all of the DH and DL characteristics (other than coverage) are \textit{monotonic}, which means that if a new configuration $c$  is added to a set of configurations $\mathcal{S}$, then $\mathcal{S} \cup \{c\}$ cannot have larger DH characteristics nor smaller DL characteristics than those of $\mathcal{S}$.
For example, adding a configuration does not decrease the total deployment time, so this is a monotone DL characteristic.
Note that coverage as defined here is always monotone.

Let $\mathcal{A}_i$ be all subsets of size $i$, and suppose all non-coverage characteristics are monotone.
Let $a_i$ be a subset in $\mathcal{A}_i$, and let $c$ be any configuration not in $a_i$.
If the coverage of $a_i \cup \{c\}$ is more than $a_i$, then we need to observe some subset in $\mathcal{A}_{i+1}$ (because $a_i \cup \{c\}$ is one such subset).
However, if the coverage does not increase for any subset in $\mathcal{A}_i$ with any new configuration $c$, then we can terminate the algorithm because (1) the coverage does not increase, and (2) the characteristics are monotone.
We give a more detailed description in Algorithm~\ref{alg:pareto_frontier}.
In practice, all of the characteristics we have used are monotone, and the algorithm benefits because most configurations in the Pareto frontier had fewer than four configurations, a significant improvement over the brute-force strategy.
We present and discuss various points of Pareto frontiers derived from this algorithm in Section~\ref{sec-evaluation}.

An advantage of our algorithm is that it is highly likely that a configuration will cover the
artifacts employed by any observed sample.  
The construction of Algorithm~\ref{alg:pareto_frontier} produces minimal subsets of configurations 
(i.e., deleting any configuration from any subset will cause coverage to decrease).
Indeed, as demonstrated in Section~\ref{sec-evaluation}, most of the points found on the Pareto frontier involved a very small number of configurations. 

\begin{algorithm}
	Generate the covering $C$ with rows $R$ as configurations, and columns as (monotone) characteristics.
	
	PreviousCoverage $\gets$ $\emptyset$.
	PointsToConsider $\gets$ $\emptyset$.
	
	\For{$i=1$ to $|R|$} {
	    NewCoverage $\gets$ $\emptyset$.
	    
	    \For{each subset $S$ of size $i$ of $R$}{
	        Add the coverage of $S$ to NewCoverage, and both the coverage and costs of $S$ to PointsToConsider.
	        
	        Call the \textit{parent} of $S$ to be every subset of $S$ of size $|S|$-1 (i.e., deletion of a single element).
	    }
	    
	    \If{the coverage of each subset in NewCoverage is the same as its parents in PreviousCoverage}{
	        Exit this loop.
	    }
	    \Else{
	        PreviousCoverage $\gets$ NewCoverage.
	    }
	}

    Output the Pareto frontier of PointsToConsider using non-dominated sorting.

    \caption{Pareto Generation of Configurations via Coverings when the characteristics are monotone.\label{alg:pareto_frontier}}
\end{algorithm}

\section{Empirical Evaluation}
\label{sec-evaluation}

\thename{} adapts coverings to choose artifact mitigation
strategies that enable the accurate and rapid analysis of 
stealthy malware that would otherwise take significant effort to analyze and
understand.  In this section, we present results from two empirical evaluations
of \thename{}.

We begin by introducing an indicative use case (see
Section~\ref{sec-workflow}). Consider an enterprise that desires to use a 
set of servers with finite capacity for automated malware classification and
triage.  
We assume that low-latency analysis of stealthy samples is paramount: 
given a fixed set of computing resources, we want the analysis of a given
sample to complete as quickly as possible (e.g., to support subsequent
human analysis, defense creation, signature generation, etc.). 
We further assume that the input samples are stealthy, and the analysis tool must mitigate
the artifacts exposed to each sample to prevent subversion.  Although it might be
possible to use all servers available to the enterprise to mitigate all
potential artifacts, this is not an efficient use of resources and does
not provide the lowest analysis latency.
Instead, we apply our algorithm to determine which sets
of artifacts are to be mitigated by each server. This minimizes
the latency of analyzing each sample across all available servers
while maximizing the combined analysis power of all available servers. 


To evaluate our approach, we consider three research questions:

\begin{description}

\item[RQ1] \textbf{Coverage} --- Does \thename{} produce artifact mitigation
configurations that effectively covers stealthy malware samples?

\item[RQ2] \textbf{Scalability} --- Does  \thename{} produce artifact mitigation
configurations that admit low-cost, high-throughput automated stealthy malware analyses?

\item[RQ3] \textbf{Efficiency} --- What tradeoffs exist in the resource costs and coverage space among the configuration sets produced by \thename{}?
\end{description}

We first discuss the corpus of malware we used in our evaluation.  Then, we discuss each research question in turn. 

\subsection{Malware Corpus Selection}

\label{sec-malware-selection}

We consider stealthy malware that targets Windows.  Of the many available malware corpora,
only a few focus directly on stealthy malware, in part because they are so 
difficult to analyze automatically.  We studied two of these in detail (BareCloud~\cite{kirat2014barecloud} and an anonymous
security company) and found that they the labels were inadequate for our purpose because they did not label the specific artifacts used by each
sample.  A sample might be labeled ``device id detection,'' for example, rather than
listing the specific device it checked for.  
Instead, we obtained a set of 1535 unique samples from independent security researchers, which are analyzed according to the artifacts
they use.
This dataset consists only malware samples that have been manually identified as stealthy and curated precisely.
Other work has used larger malware databases for similar experiments~\cite{bayer2009scalable,cheng2018towards}, but as mentioned above these datasets are not labeled with enough specificity for our study.

\begin{figure}
\centering
\includegraphics[width=\columnwidth]{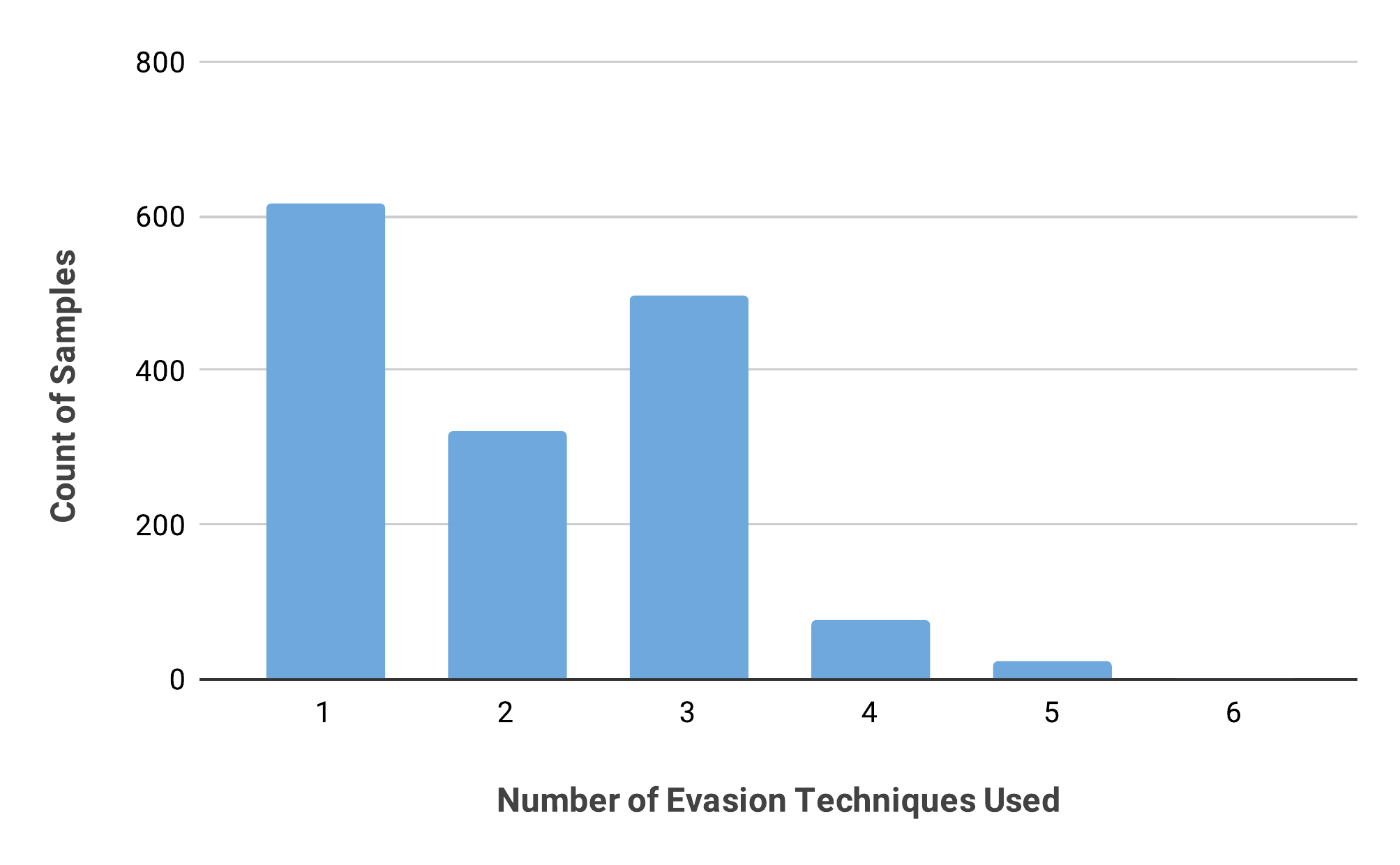}

\caption{
Distribution of malware samples in our dataset according to the number of unique artifacts employed. For example, more than 600 of our 1535 samples employed a single artifact.  The graph is not cumulative.  
\label{fig:freq_dist}}

\end{figure}

\begin{figure}
\centering
\includegraphics[width=\columnwidth]{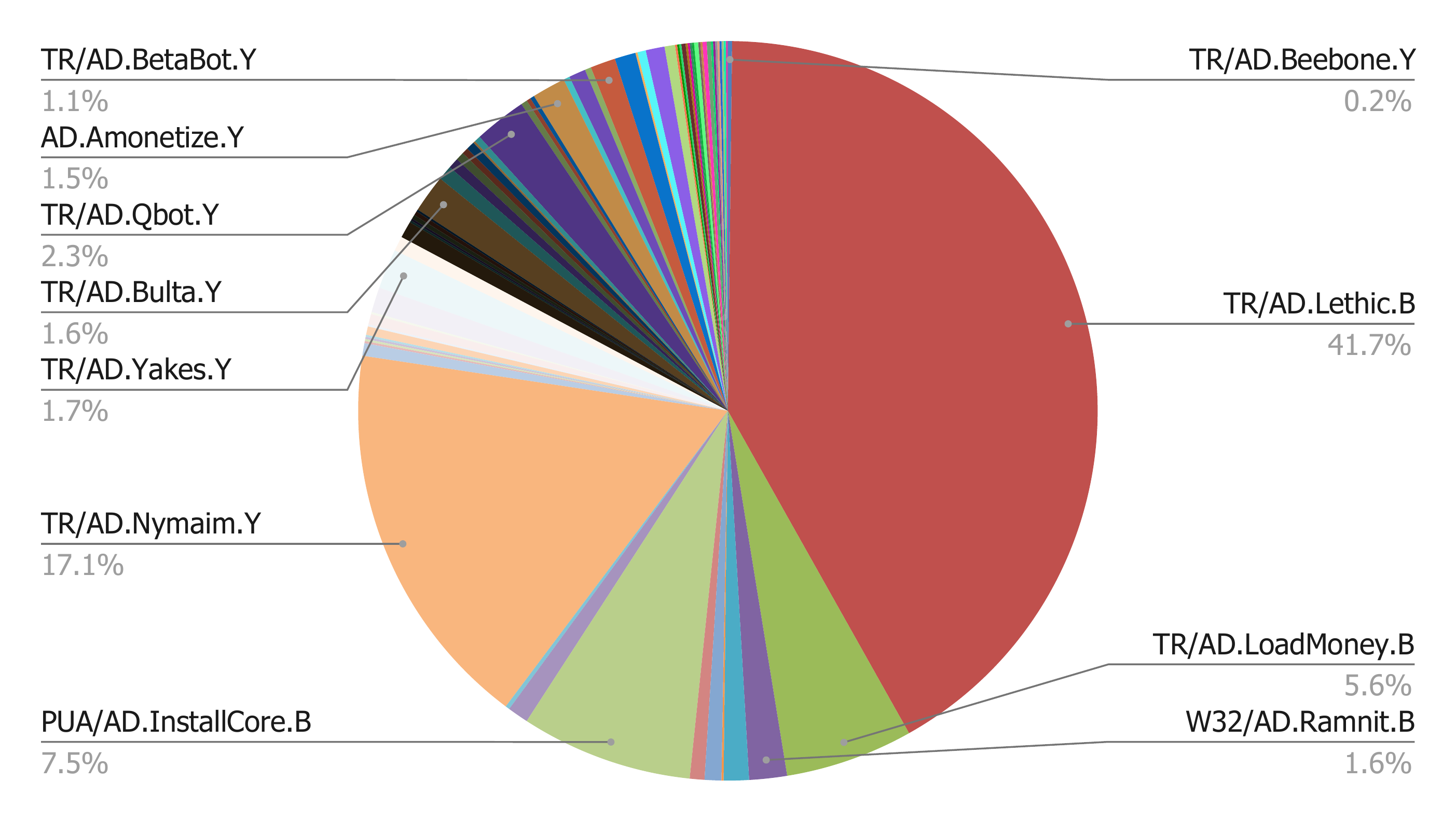}

\caption{
Taxonomy of malware samples contained in the corpus, which includes samples of several popular families such as
Lethic (Trojan), Nymaim (Trojan), and InstallCore (Potentially Unwanted Program (PUA)). 
\label{fig:taxonomy}}

\end{figure}

Figure~\ref{fig:freq_dist} shows that each individual malware sample in our corpus uses 
between one and five evasion techniques, thus confirming our hypothesis that most malware considers only a few artifacts and supporting our design decisions for \thename{}.
In addition, we show a taxonomy of malware families in our corpus in Figure~\ref{fig:taxonomy}.


\begin{figure}
\center
\includegraphics[width=\columnwidth]{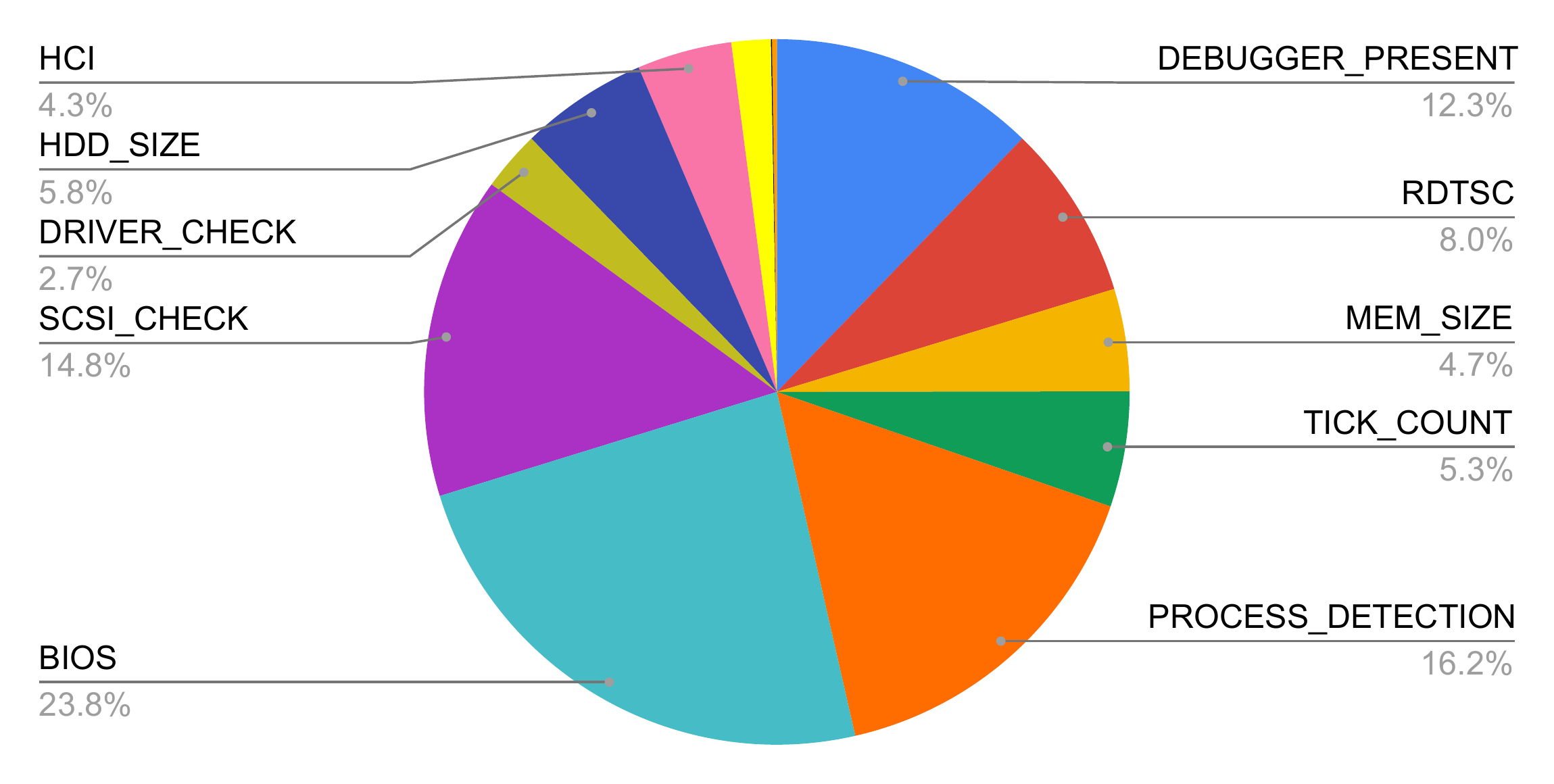}

\caption{Frequency of artifacts detected by samples in the malware dataset.\label{fig:artifacts_freq}}

\end{figure}

\begin{figure}
\center
\includegraphics[width=\columnwidth]{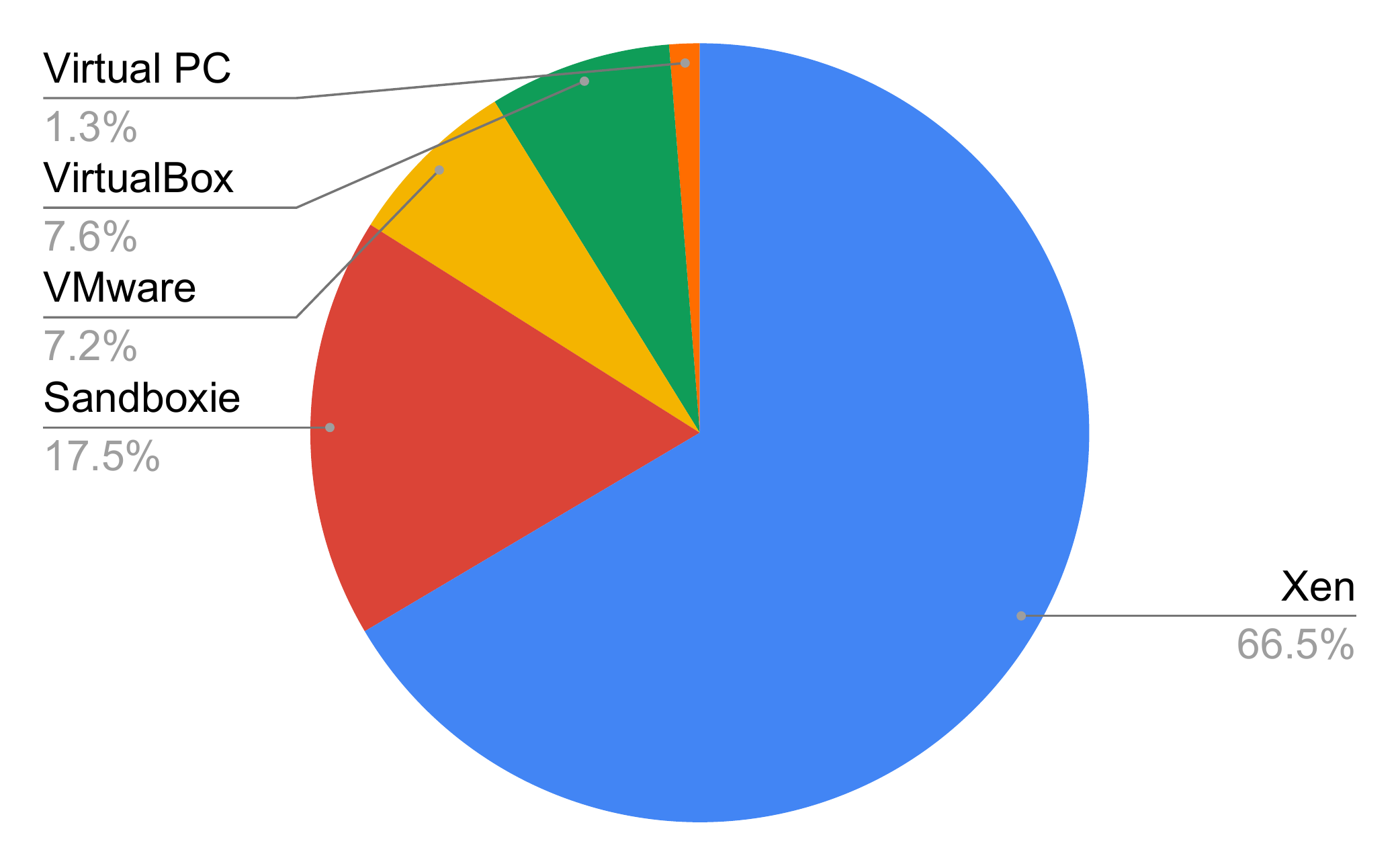}

\caption{Distribution of process names checked for by stealthy malware. \label{fig:process_freq}}

\end{figure}

We categorized the samples based on the artifacts they are looking for in the system, summarized in Figure~\ref{fig:artifacts_freq}.   
Among these artifacts, checking for BIOS and SCSI device metadata were common.  Additionally, many of our samples checked for the existence of specific processes (e.g., helper programs for in-guest clipboard access, video acceleration, etc.).
We categorized which specific process was used by each stealthy malware sample, shown in Figure~\ref{fig:process_freq}.
In particular, Xen service (\texttt{xenservice.exe}) is the most frequently-checked process among other processes used.

\subsubsection{Pafish}

In addition, we used pafish~\cite{pafish}, an open source tool that enumerates
common checks used by stealthy malware, to determine whether a given
configuration could provide coverage over specific artifacts.  Pafish is well-suited to this task because it can be configured to
check or ignore specific artifacts.
We used pafish to confirm the sets of artifacts mitigated by each configuration before we applied each configuration to malware samples in our dataset.


\begin{figure*}
\centering
\includegraphics[width=.95\textwidth]{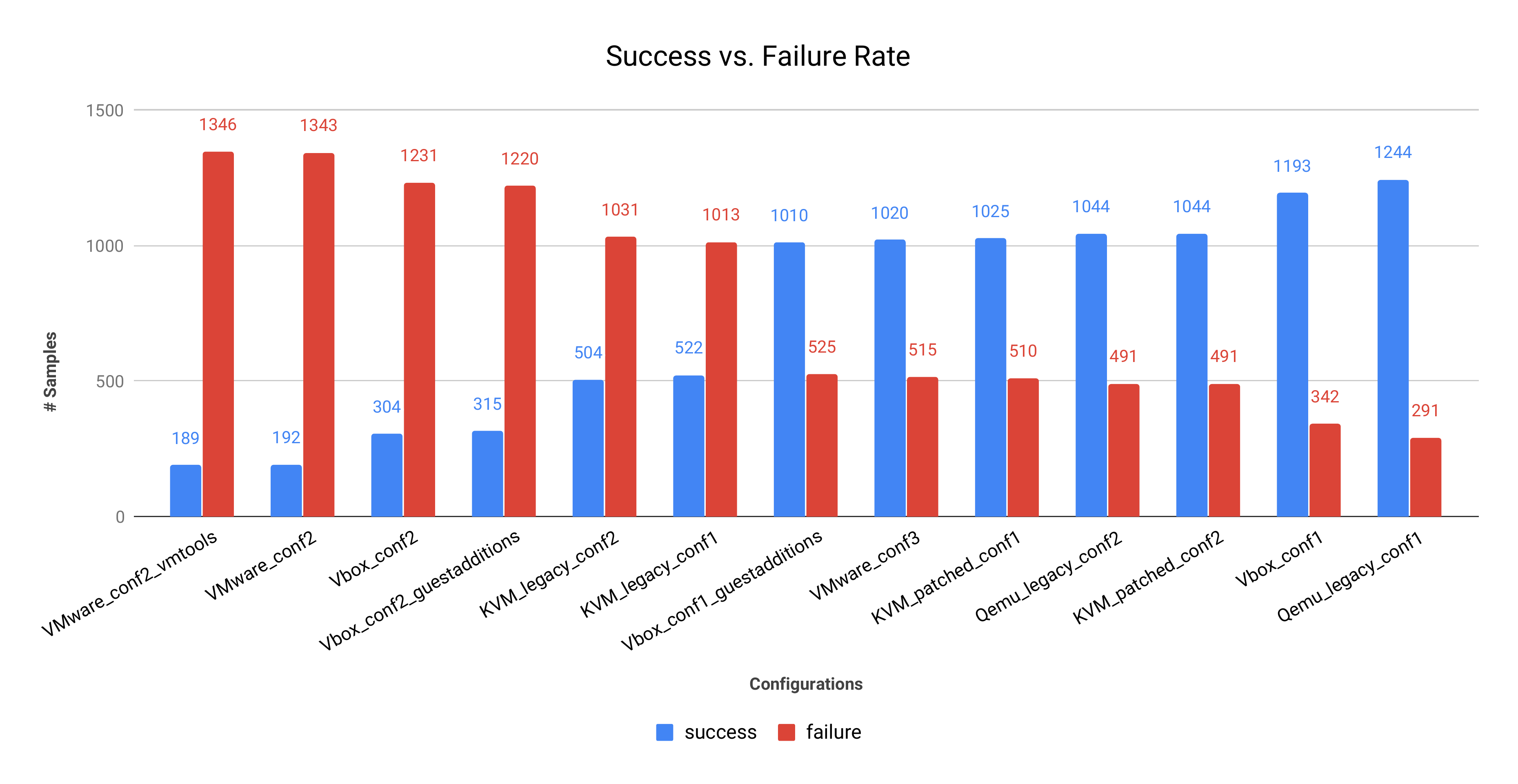}
\caption{Success and failure counts for each tested configuration, when run against the 1535 stealthy malware samples. \label{fig:detection_rate}}

\end{figure*}

\subsection{RQ1: Coverage --- Artifact Mitigation}

\label{sec:artifact_mitigation}

In this experiment, \thename{} assigns artifact mitigation strategies 
to analysis servers. 
We say that the \emph{configuration set size} is the number of configurations combined together --- this is an input parameter that represents the number of distinct configurations that the user is willing to run concurrently.
For example, if more servers are available for analysis, a larger configuration set size can be selected. 
We say that a stealthy malware sample is successfully
analyzed if at least one configuration in the configuration set produced by \thename{} mitigates all of the artifacts it uses.

\begin{table*}
\centering

    \caption{List of tested configurations and the artifacts they mitigate.  Each column corresponds to whether a specific category of artifact is mitigated in that configuration.   Note that some configurations support different backend (e.g., qemu\_legacy\_conf1 can be run in both KVM and QEMU), yielding differing artifact mitigations.\label{tab-config}}

\begin{small}

\resizebox{\textwidth}{!}{\setlength\tabcolsep{2pt}\begin{tabular}{cccccccccccccccccc} 
\toprule
\textbf{Index}    & \textbf{Backend}      & \textbf{Configuration} & \textbf{Process} & \textbf{Debugger} & \textbf{CPUID} & \textbf{RDTSC} & \textbf{CPU \#} & \textbf{Invalid Inst.} & \textbf{TickCount} & \textbf{HCI} & \textbf{BIOS} &  \textbf{File Check} & \textbf{HDD - SCSI} & \textbf{Disk size} & \textbf{Memory} & \textbf{MAC} & \textbf{ACPI}\\  \midrule
 1     & \multirow{2}{*}{KVM}           
        & qemu\_patched\_conf1           & \checkmark & \checkmark & -- & -- & \checkmark & \checkmark & -- & \checkmark & \checkmark & -- & -- & -- & -- & \checkmark & \checkmark \\
2   &   & qemu\_patched\_conf2           & \checkmark & \checkmark & -- & -- & \checkmark & \checkmark & -- & \checkmark & \checkmark & -- & \checkmark & -- & \checkmark & -- & \checkmark \\ \midrule
3     & \multirow{3}{*}{VMWare}                  
        & vmware\_conf3                  & \checkmark & \checkmark & -- & -- & \checkmark & \checkmark & -- & \checkmark & \checkmark & -- & -- & -- & -- & \checkmark & \checkmark \\
4   &   & vmware\_conf2                  & \checkmark & \checkmark & -- & -- & -- & -- & -- & \checkmark & -- & \checkmark & -- & -- & \checkmark & -- & \checkmark \\ 
5   &   & vmware\_conf2\_vmtools         & -- & \checkmark & -- & -- & \checkmark & -- & -- & \checkmark & -- & -- & -- & -- & \checkmark & -- & \checkmark \\ \midrule
6     & \multirow{2}{*}{KVM}                  
        & qemu\_legacy\_conf1            & \checkmark & \checkmark & -- & -- & \checkmark & \checkmark & -- & \checkmark & \checkmark & \checkmark & -- & -- & \checkmark & \checkmark & \checkmark \\
7   &   & qemu\_legacy\_conf2            & \checkmark & \checkmark & -- & -- & -- & \checkmark & -- & \checkmark & \checkmark & \checkmark & -- & -- & -- & -- & \checkmark \\ \midrule
8     & \multirow{4}{*}{Virtualbox}           
        & vbox\_conf1\_guestadditions    & -- & \checkmark & -- & -- & -- & \checkmark & -- & \checkmark & \checkmark & -- & \checkmark & \checkmark & -- & -- & -- \\
9   &   & vbox\_conf2\_guestadditions    & -- & \checkmark & -- & -- & \checkmark & \checkmark & -- & \checkmark & -- & -- & \checkmark & \checkmark & -- & \checkmark & \checkmark \\
10   &   & vbox\_conf1                   & \checkmark & \checkmark & \checkmark & -- & \checkmark & \checkmark & -- & \checkmark & \checkmark & \checkmark & \checkmark & \checkmark & \checkmark & -- & -- \\
11   &   & vbox\_conf2                   & \checkmark & \checkmark & -- & -- & -- & \checkmark & -- & \checkmark & -- & \checkmark & \checkmark & \checkmark & -- & \checkmark & -- \\ \midrule
12     & \multirow{2}{*}{QEMU}           
         & qemu\_legacy\_conf1           & \checkmark & \checkmark & \checkmark & \checkmark & \checkmark & \checkmark & \checkmark & \checkmark & \checkmark & \checkmark & \checkmark & -- & -- & \checkmark & \checkmark \\
13   &   & qemu\_legacy\_conf2           & \checkmark & \checkmark & \checkmark & -- & \checkmark & \checkmark & -- & \checkmark & \checkmark & \checkmark & \checkmark & -- & \checkmark & -- & \checkmark \\ 
\bottomrule
\end{tabular}}
\end{small}
\end{table*}

\label{sec-cic}



For each configuration, we represent artifact coverage as a
bit-array in which each set bit implies that that particular artifact has been
successfully mitigated in the environment. Table~\ref{tab-config} gives
details about each configuration instance.

We use VMWare, VirtualBox, KVM, and QEMU backends for virtualizing guests to complete an analysis of each sample.
We use 13 different configurations across each of these backends for conducting analyses.
Each configuration implements a subset of mitigations against each class of artifacts.
For example, the \emph{qemu\_patched\_conf1} contains intentionally low RAM size ($<1$GB), exposing the RAM detection family of artifacts, but also contains custom patches that remove all QEMU-related hardcoded strings throughout the source.
In contrast, the \emph{VMWare\_conf2} configuration employs the VMWare Tools suite for faster execution, exposing process names (i.e., of VMWare Tools).
Broadly, we designed and implemented these configurations by considering the families of artifacts exposed by our dataset (Section~\ref{sec-malware-selection}) and the expected complexity in mitigating each artifact family across each virtualization backend.

%
%
%
%

We compute whether at least one configuration covers each evasive sample by analyzing traces of API call invocations, including arguments passed to each call and the corresponding output.
We developed a module (``Detox'' engine in Figure~\ref{fig:workflow}) to wrap and unify multiple Virtual Machine Introspection (VMI) APIs, including Icebox~\cite{amiauxicebox}, PyReBox, DRAKVUF~\cite{lengyel2014scalability}, and VMWare VProbes~\cite{westphal2014vmi}.  
Thus, we collect multiple API trace logs for each sample for each configuration, based on the virtualization backend used. 
Next, we aggregate these API trace logs to bridge the semantic gap~\cite{saberi2014hybrid, jain2014sok}: doing so allows reconstructing higher abstraction API traces invoked against the guest OS. 

Given each trace of each sample, we confirm detection results based on the malware's behavior across all configurations.
Specifically, we follow the malware execution trace up to the point when it starts to create, manipulate, or remove a memory section, segment, or page using APIs such as \texttt{NtCreateSection}, \texttt{NtMapViewOfSection}, or \texttt{NtSetContext}.
Then, we compare these results against ground truth established in our corpus to ensure that the malicious process executed completely.
If the analysis differs from the ground truth (e.g., if the sample detects the environment and hides its behavior), we say that configuration does \emph{not} cover the sample.
If there exists at least one configuration that \emph{does} cover the sample, we call that sample covered. 
We show the detection rate of each configuration across our entire corpus of malware in Figure~\ref{fig:detection_rate}.

\begin{figure}[h]
    \centering
    \resizebox{\columnwidth}{!}{\input{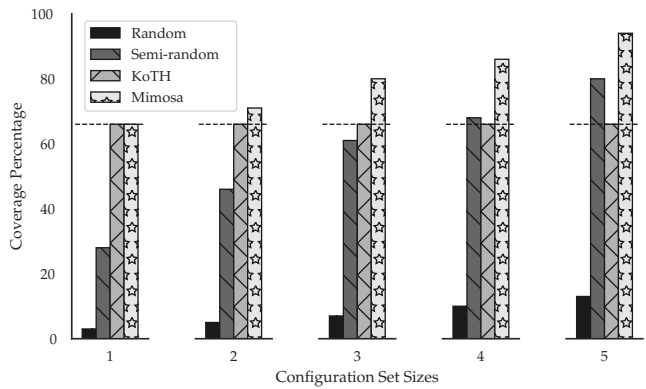}}
    \caption{
    Proportion of stealthy malware samples covered for different configuration set sizes for various techniques.
    Random indicates a randomly-generated coverage vector.
    Semi-random represents a randomly-selected subset of our 13 configurations.
    King-of-the-Hill represents the best single configuration from our set of 13.
    Our approach achieves higher levels of coverage compared to the best available single configuration.
    \label{fig:coverage_comparison}}
\end{figure}

\subsubsection{RQ1 Result Summary}
\label{sec-rq1-summary}


Our mitigation strategies and corresponding configurations provide
varying coverage levels across an indicative dataset of 1535 stealthy
malware samples, allowing us to explore the trade-off space between coverage
provided by analysis tools and the cost of deploying those tools or acquiring
analyses.  In Figure~\ref{fig:coverage_comparison}, we show a the level of coverage achieved by our approach compared to 
other approaches versus configuration set size.
Specifically, we measure the coverage achieved by a set of configurations of a specific size for
(1) Random --- a randomly-generated coverage vector,
(2) Semi-random --- a randomly-selected configuration from our set of 13 configurations (shown in Table~\ref{tab-config}),
(3) King-of-the-Hill (KoTH) --- the best single configuration from our set of 13 configurations, and 
(4) \thename{}, the set of configurations selected by our approach.
For this set of experiments, we averaged 10 trials.

We view KoTH as the baseline for automated malware analysis systems that do not use our approach (e.g., companies that pick the ``best'' sandbox they can, and scale it up to multiple machines in a cluster). 
Our approach achieves 97\% coverage when combining five configurations, compared to KoTH, which achieves 65\% coverage.
This suggests our approach can generate configurations of malware analysis environments that can apply to most stealthy malware samples in an indicative corpus.

\subsection{RQ2: Scalability --- Automated Analysis}

We also evaluated our approach with respect to malware analysis throughput.  
Because our approach also considers the relative costs (e.g., overhead, disk utilization)
of each configuration, we can measure our system's effectiveness at scale. 
For example, if a given configuration \emph{does not} cover a given sample, that configuration wastes time and resources attempting to execute that sample. 
Thus, we can compute the amount of resources \emph{wasted} by considering the total resources consumed by configurations executing samples that were not covered by those configurations.

We measure the time wasted by a configuration using virtual machine introspection (VMI) to reconstruct events that occur within each configuration guest environment from low-level execution traces collected for each sample. 
We compared these execution traces against ground truth execution traces gathered for each sample (provided as part of our malware dataset). 
Each sample's collected and ground truth traces were compared using the trace merging algorithm introduced by Virtuoso~\cite{dolan2011virtuoso}, VMWatcher~\cite{jiang}, and VMWare VProbes~\cite{westphal2014vmi}.
For each sample in each configuration, we report the time $t$ at which the measured and ground truth traces diverged --- where a sample's anti-analysis technique caused the execution to differ from the ground truth.
If the traces never diverge, then we conclude the sample was covered.
Thus, for each uncovered sample and configuration, we report the time wasted as the difference between time $t$ and some maximum timeout (configured as 2s here; state-of-the-art typically uses 5s timeouts~\cite{lengyel2014scalability}).

\begin{figure}[h]
    \centering
    \resizebox{\columnwidth}{!}{\input{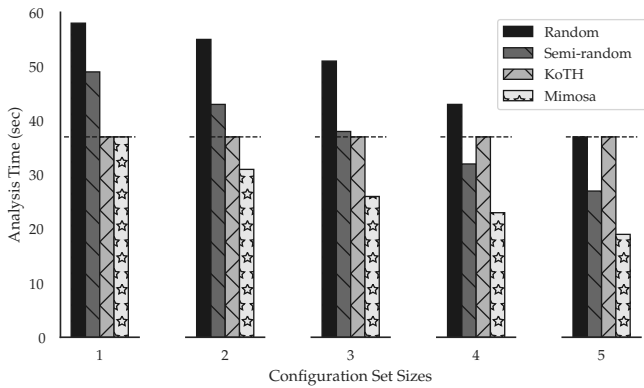}}

    \caption{
    Average analysis time wasted executing each sample for sets of different configuration sizes. 
    Random refers to a randomly-generated coverage vector.  
    Semi-random refers to a random subsets of our 13 configurations.
    King-of-the-Hill represents the best single configuration selected from our 13 configurations.
    Our approach wastes the least amount of time failing to execute stealthy samples, enabling higher automated analysis throughput.
     \label{fig:idle_time}}

\end{figure}

Figure~\ref{fig:idle_time} shows a comparison of time wasted of various approaches versus configuration set size, as described in Section~\ref{sec-rq1-summary}.
Our approach spends 3X less CPU time executing samples that \emph{are not covered} by configurations.
As a result, our approach can scale analysis of malware samples 3X over state-of-the-art by accurately analyzing a higher proportion of samples in less time.

\subsection{RQ3: Efficiency --- Analysis Tradeoffs}

In this section, we consider tradeoffs between analysis resource cost and stealthy malware sample coverage. 
Recall that a stealthy malware sample is \emph{covered} if all of the artifacts it uses are mitigated.
We evaluate coverage with respect to two cost functions: memory utilization and disk throughput. Both are relevant for scalable automated malware analysis.

We analyzed each of the 1535 stealthy malware samples.
For each sample, we determined which set of configurations would mitigate the artifacts used by that sample, then measured how much of a resource was used during that sample's execution.
In particular, we measured disk throughput (bytes per second) and memory utilization (approximated by measuring average free bytes during execution).
We used \thename{} to generate a Pareto front by considering which subsets of configurations would require which levels of resource to achieve a particular degree of coverage.







Table~\ref{tab:cost-config} shows the Pareto front and indicative points for the memory utilization cost function.  As an example, the point with the highest coverage (i.e., 1432 out of 1535 samples analyzed successfully) required an average of 718MB during execution, while the configurations with lower coverage (e.g., 1020 samples) used only 316MB of memory.  Overall, this graph shows how to balance the tradeoff between 
malware analysis tool configurations with respect to memory usage.  

Similarly, 
Table~\ref{tab:disk-config} show the Pareto front for the disk throughput cost function.  As before, there is a tradeoff  between how many samples are covered and the disk usage is required to obtain analyses per sample.

\subsection{RQ3 Tradeoffs Summary}

\thename{} enables finding a Pareto-optimal point that provides accurate stealthy malware analyses while minimizing the resource allocation required to obtain those analyses. 

%
%

\begin{table}
\centering
\caption{Indicative points in the Pareto front comparing samples covered with memory utilization. \label{tab:cost-config}}

\begin{small}
\setlength\tabcolsep{1pt}\begin{tabular}{lcc} 
\toprule
&                      \textbf{Samples}  & \textbf{Avg. Available} \\ 
\textbf{Configuration Set} & \textbf{Covered} &\textbf{Memory(MB)} \\\midrule
vbox\_conf2 & \multirow{9}{*}{1432} & \multirow{9}{*}{718}\\ 
vbox\_conf1 & & \\
qemu\_legacy\_conf1 & & \\
qemu\_patched\_conf1 & & \\
vmware\_conf3 & & \\
qemu\_legacy\_conf2 & & \\
qemu\_legacy\_conf1 & & \\
vbox\_conf2\_guestadditions & & \\  
vbox\_conf1\_guestadditions & & \\ \midrule
vmware\_conf3 & 1020 & 316 \\
\bottomrule
\end{tabular}
\end{small}
\end{table}

%

\begin{table}[t]
\centering
\caption{Indicative points in the Pareto front comparing samples covered with disk throughput.\label{tab:disk-config}}
\begin{small}
\setlength\tabcolsep{1pt}\begin{tabular}{lcc} 
\toprule
&                      \textbf{Samples}  & \textbf{Avg. Disk Write} \\ 
\textbf{Config Set} & \textbf{Covered} &\textbf{(KBytes/sec)} \\\midrule
vbox\_conf1 & \multirow{3}{*}{1432} & \multirow{3}{*}{397}\\ 
qemu\_legacy\_conf1 & & \\
qemu\_legacy\_conf2 & & \\ \midrule
qemu\_legacy\_conf2 & 1044 & 209 \\
\bottomrule
\end{tabular}
\end{small}
\end{table}

%
%

%
%
%

%
%

\section{Discussion}

In this section, we discuss (1) potential threats to the validity of the
experimental results, (2) using \thename{} 
for controlling an adaptive malware analysis system, and (3) potential
future improvements that can be made to cost functions.

\subsection{Threats to Validity}

First, we characterized artifact families according to conceptual similarity.
The artifact families ultimately inform what structure the corresponding
covering takes.  There is no standard method for classifying artifacts in this
manner---the effectiveness or utility of \thename{} could change depending on
the specific assumptions we made about which artifacts are categorically similar. 

Second, our experimental approach for RQ2 measured execution time only while the
sample was actively executing.  In practice, there are other considerations
that have impact on the overall efficiency of malware analysis (e.g.,
restoring clean virtual disks, reloading the OS image, etc.).  

Third, although our evaluation incorporated 1535 stealthy malware samples from the
wild, we produced configurations whose costs were measured in isolation
(e.g., we measured CPU utilization separately from memory utilization).  Additional
engineering
effort is required to construct a production-quality end-to-end system that uses the
configurations produced by \thename{} to apply to a real set of hardware. 

\subsection{Remarks on Adaptability}

\thename{} takes as input a set of modeled mitigation strategies and
associated costs, and it produces as output a coverage-optimal,
low-heuristic-cost array of strategies.   This approach can be
extended to adapt over time to changes in the distribution of stealthy malware.  For
example, if new artifacts are discovered or if the costs associated with
mitigating each one changes with technology, our overall approach and algorithms will still be
applicable as a tool for finding cost-optimal analysis configurations.  

As a specific example, recent work leveraged ``wear and tear'' of virtual
machine environments~\cite{miramirkhani2017spotless}.  In essence, malware
samples can look for evidence that an environment is ``aged.'' An analyst that
spins up a vanilla VM image may fall victim to a sample that detects if the
environment is pristine and newly-created.  That is, the perceived ``age'' of
the virtualized environment is the artifact.  Malware campaigns like Dyre and
Dridex use heuristics like (1) investigating the clipboard for evidence of
random strings associated with normal use, and (2) registry keys to track
historical use of common prorgrams (e.g., Microsoft Word).
We do not
include such artifacts in our prototype coverage calculation because our dataset did not contain samples that exploited wear and tear artifacts; however, they can be readily incorporated by
implementing a corresponding mitigation.  
For example, our prototype currently moves random files to the Desktop, Recycle bin, and Temp directories, and it also injects decoy entries in the Registry. 
We could introduce this as a full mitigation in our framework:
the coverings vector would be augmented to reflect this new artifact so
that it is covered in the optimally-generated configurations.

\subsection{Remarks on Cost Functions}

\thename{} currently considers optimizing for cost, which can be captured in several ways: CPU utilization, memory utilization, and runtime
overhead with respect to latency.  However, these one-dimensional approaches may
admit coverings that are difficult to interpret.  For example, in a cluster of
10 servers, assigning nine servers to do no mitigation (minimal cost) and one
server to run bare metal (maximal coverage) is a well-formed solution.

We also discussed a second parameter that captures \emph{benefit}: coverage
of stealthy malware samples is important for acquiring faithful, interpretable
analyses.  For example, if we know a mitigation strategy will cover 90\% of
stealthy malware, we may be willing to pay a higher cost to use that strategy
because of its overall coverage.   On the other hand, a strategy that only
covers 2\% of stealthy malware in the wild may be disregarded.  While we
examined the cost-benefit space in our evaluation, future work will include
a multidimensional heuristic search to find optimal coverings with respect to
more complex cost functions.

\section{Related Work}
\label{sec-related} 

Various projects have focused on detecting and evading analysis
systems in both x86
executables~\cite{anti-emulator-ISC07,chen2008towards,redpill,quist2006detecting}
and mobile devices (e.g., Android~\cite{Jing:2014:MAG:2664243.2664250}).
In this section, we discuss this work in three categories:  (1) malware
detection using behavioral analysis, (2) malware analysis using virtual machine
infrastructure, and (3) malware analysis using bare-metal machines.

%

\subsection{Stealthy Malware Detection} 

Current stealthy malware analysis techniques generally rely either on human
creativity (e.g., debugging with IDA Pro~\cite{idapro} or
OllyDbg~\cite{ollydbg}) or heavy-weight analysis tools that incur significant
overhead (e.g., MalT~\cite{malt-sp15} or Ether~\cite{ether}).  Moreover,
differencing approaches, such that of as Balzarotti \emph{et
al.}~\cite{balzarotti2010efficient}, work by executing a sample in multiple
instrumented environments and use the difference in runs to determine which
artifact is used by the sample, potentially wasting resources.

Balzarotti \emph{et
al.}~\cite{balzarotti2010efficient} demonstrate the ability to detect evasive
behaviors by running malware in various runtime environments and comparing
their system calls.
Lindorfer \emph{et al.}~\cite{10.1007/978-3-642-23644-0_18} later employed a
similar technique, but used various malware sandboxes and scored their evasive
behaviors.  HASTEN~\cite{kolbitsch2011power} specifically focuses on stalling
malware, which is a particularly difficult evasion technique to analyze because
the malware appears benign for an extended period of time.
TriggerScope~\cite{fratantonio16:triggerscope} similarly examines Android
programs which mask there malicious behavior until a certain \emph{trigger} is
observed.   
Our technique leverages a combination of multiple environments that separately mitigate different artifact families, instead providing environments that are more likely for the sample to execute faithfully.

Our approach is conceptually related to SLIME~\cite{slime}, an automated tool
for disarming anti-sandboxing techniques employed by stealthy malware.  SLIME
runs a sample many times, each time configuring the environment to explicitly
expose certain artifacts to the sample.  In contrast, our approach seeks to
minimize the total cost of execution (or the 
resources consumed) to either identify or analyze the sample under
test.  In addition, we introduce a novel structure called a covering that helps
identify the optimal configuration for an analysis system. 

\subsection{Virtual Machine Analysis} Ether~\cite{ether} is a malware analysis
framework based on hardware virtualization extensions (e.g., Intel VT).  It
runs outside of the guest operating systems, in the hypervisor, by
relying on underlying hardware features.  BitBlaze~\cite{bitblaze} and
Anubis~\cite{anubis} are QEMU-based malware analysis systems.  They focus on
understanding malware behavior, instead of achieving better transparency.
V2E~\cite{v2e-vee12} combines both hardware virtualization and software
emulation.  HyperDbg~\cite{hyperdbg:ase10} uses the hardware virtualization
that allows the late launching of VMX modes to install a virtual machine
monitor, and run the analysis code in the VMX root mode.
SPIDER~\cite{deng2013spider} uses Extended Page Tables to implement invisible
breakpoints and hardware virtualization to hide its side-effects.
DRAKVUF~\cite{lengyel2014scalability} is another VMI-based system capable of
both user and kernel-level analysis.

We note that recent work has investigated changes to the sandboxing environment
to give it the appearance of age or use~\cite{miramirkhani2017spotless}.
For example, a dearth of
Documents, Downloads, event logs, or installed software could be a hint that
the sample is not executing in a real, vulnerable environment.  
Although our current prototype does not
address samples exhibiting such ``age'' checks, as discussed above, we could readily incorporate
it.  As with other new or yet-undiscovered artifacts, our overall framework would not change.  One would simply implement a configurable mitigation against
that new artifact and include it as a strategy used
by our coverings algorithm.

\subsection{Bare-metal Analysis} BareBox~\cite{barebox-acsac11} is a malware
analysis framework based on a bare-metal machine without any virtualization or
emulation techniques, which is used for analyzing user mode malware.  Follow up
work, BareCloud~\cite{kirat2014barecloud}, uses mostly un-instrumented
bare-metal machines, and is capable of analyzing stealthy malware by detecting
file system changes.  Willems \emph{et al.}~\cite{willems2012down} propose a
method for using branch tracing, implemented on a physical CPU, to analyze
stealthy malware.  LO-PHI~\cite{lophi:ndss16} is a system capable of both live
memory and disk introspection on bare-metal machines, which can be used for
analyzing stealthy malware.  MalT~\cite{malt-sp15} uses System Management Mode
to instrument a bare-metal system at the instruction level, exposing very few
artifacts to the system.  While LO-PHI and MalT both have high deployment
overheads, they also expose very few artifacts to samples under test; thus, either
could conceptually serve as our highest coverage (and highest cost) configuration.

\section{Conclusion}

Stealthy and obfuscated malware is expanding rapidly.  As the
security arms race continues, malware authors use increasingly sophisticated
techniques to subvert analysis.  The large volume of new malware released every year makes automated analysis increasingly mandatory to identify and understand new malware samples. Techniques to
address the volume of stealthy malware are critical.

In this paper, we introduced coverings, a novel way of representing the
problem of analyzing stealthy malware efficiently, and a prototype implementation called
\thename{}.  We studied a broad set of artifacts exposed by analysis environments and
the mitigation strategies required to prevent malware samples from using
those artifacts to subvert detection.  We modeled the mitigations using a partially ordered structure according to the number of artifacts mitigated and the cost associated with deploying that strategy.  We developed 32 such mitigation strategies.  We
presented an algorithm that finds the lowest-cost selection of mitigation
strategies to implement while guaranteeing a total coverage of the artifacts.
Finally, we empirically evaluated \thename{} using 1535 stealthy malware
samples from the wild.  We found that  \thename{} can find mitigation
strategies that reduce the overhead and memory utilization associated with
mitigating all artifacts considered. 

\section{Acknowledgments}

We thank Giovanni Vigna, Christopher Kruegel, and Hojjat Aghakhani for graciously providing the well-labeled corpus of evasive malware samples used in our evaluation.  This work would not be possible without the tireless engineering effort invested to construct such a dataset, so we are grateful for members of the community who share data to improve the state-of-the-art. 

We also thank the anonymous reviewers for their valuable comments and suggestions, and the Avira company, Alexander Vukcevic, Director of Protection Labs and QA, and Shahab Hamzeloofard for helping us with determining provenance of our malware samples. 

We gratefully acknowledge the partial support of NSF (CCF
1908633, 1763674), DARPA (FA8750-19C-0003, N6600120C4020), AFRL
(FA8750-19-1-0501), and the Santa Fe Institute. Any opinions, findings, and conclusions in this paper are those of the authors and do not necessarily reflect the views of our sponsors.

The opinions in the work are solely of the authors, and do not necessarily reflect those of the U.S. Army, U.S. Army Research Labs, the U.S. Military Academy, or the Department of Defense.



\bibliographystyle{IEEEtranS}
\bibliography{./bib/main}

\begin{thebibliography}{10}
\providecommand{\url}[1]{#1}
\csname url@samestyle\endcsname
\providecommand{\newblock}{\relax}
\providecommand{\bibinfo}[2]{#2}
\providecommand{\BIBentrySTDinterwordspacing}{\spaceskip=0pt\relax}
\providecommand{\BIBentryALTinterwordstretchfactor}{4}
\providecommand{\BIBentryALTinterwordspacing}{\spaceskip=\fontdimen2\font plus
\BIBentryALTinterwordstretchfactor\fontdimen3\font minus
  \fontdimen4\font\relax}
\providecommand{\BIBforeignlanguage}[2]{{%
\expandafter\ifx\csname l@#1\endcsname\relax
\typeout{** WARNING: IEEEtranS.bst: No hyphenation pattern has been}%
\typeout{** loaded for the language `#1'. Using the pattern for}%
\typeout{** the default language instead.}%
\else
\language=\csname l@#1\endcsname
\fi
#2}}
\providecommand{\BIBdecl}{\relax}
\BIBdecl

\bibitem{aspack}
``{ASPack},'' \url{http://www.aspack.com}, retrieved November 2016.

\bibitem{upx}
``{UPX: The Ultimate Packer for eXecutables},'' \url{https://upx.github.io},
  retrieved November 2016.

\bibitem{amiauxicebox}
B.~Amiaux, L.~Farey, J.-M. Borello, and T.~Rennes, ``Icebox: analyse de
  malwares par introspection de machine virtuelle.''

\bibitem{anubis}
Anubis, ``Analyzing unknown binaries,'' \url{http://anubis.iseclab.org}.

\bibitem{volatility}
\BIBentryALTinterwordspacing
M.~Auty, A.~Case, M.~Cohen, B.~Dolan-Gavitt, M.~H. Ligh, J.~Levy, and
  A.~Walters. Volatility framework - volatile memory extraction utility
  framework. [Online]. Available: \url{http://www.volatilityfoundation.org/}
\BIBentrySTDinterwordspacing

\bibitem{anti-vm-exception}
E.~Bachaalany, ``{Detect if your program is running inside a Virtual
  Machine},''
  \url{http://www.codeproject.com/Articles/9823/Detect-if-your-program-is-running-inside-a-Virtual}.

\bibitem{balzarotti2010efficient}
D.~Balzarotti, M.~Cova, C.~Karlberger, and G.~Vigna, ``Efficient detection of
  split personalities in malware.'' in \emph{Networks and Distributed Systems
  Security Symposium}, 2010.

\bibitem{bayer2009scalable}
U.~Bayer, P.~M. Comparetti, C.~Hlauschek, C.~Kruegel, and E.~Kirda, ``Scalable,
  behavior-based malware clustering.'' in \emph{NDSS}, vol.~9.\hskip 1em plus
  0.5em minus 0.4em\relax Citeseer, 2009, pp. 8--11.

\bibitem{anti-debug-bh12}
R.~Branco, G.~Barbosa, and P.~Neto, ``{Scientific but Not Academical Overview
  of Malware Anti-Debugging, Anti-Disassembly and Anti-VM Technologies},'' in
  \emph{Black Hat}, 2012.

\bibitem{bulazel2017survey}
A.~Bulazel and B.~Yener, ``A survey on automated dynamic malware analysis
  evasion and counter-evasion: Pc, mobile, and web,'' in \emph{Proceedings of
  the 1st Reversing and Offensive-oriented Trends Symposium}.\hskip 1em plus
  0.5em minus 0.4em\relax ACM, 2017, p.~2.

\bibitem{anti-debug-chen-DSN08}
X.~Chen, J.~Andersen, Z.~Mao, M.~Bailey, and J.~Nazario, ``Towards an
  understanding of anti-virtualization and anti-debugging behavior in modern
  malware,'' in \emph{Proceedings of the 38th Annual IEEE International
  Conference on Dependable Systems and Networks (DSN '08)}, 2008.

\bibitem{chen2008towards}
X.~Chen, J.~Andersen, Z.~M. Mao, M.~Bailey, and J.~Nazario, ``Towards an
  understanding of anti-virtualization and anti-debugging behavior in modern
  malware,'' in \emph{Dependable Systems and Networks With FTCS and DCC, 2008.
  DSN 2008. IEEE International Conference on}.\hskip 1em plus 0.5em minus
  0.4em\relax IEEE, 2008, pp. 177--186.

\bibitem{cheng2018towards}
\BIBentryALTinterwordspacing
B.~Cheng, J.~Ming, J.~Fu, G.~Peng, T.~Chen, X.~Zhang, and J.-Y. Marion,
  ``Towards paving the way for large-scale windows malware analysis: Generic
  binary unpacking with orders-of-magnitude performance boost,'' in
  \emph{Proceedings of the 2018 ACM SIGSAC Conference on Computer and
  Communications Security}, ser. CCS ’18.\hskip 1em plus 0.5em minus
  0.4em\relax New York, NY, USA: Association for Computing Machinery, 2018, p.
  395–411. [Online]. Available: \url{https://doi.org/10.1145/3243734.3243771}
\BIBentrySTDinterwordspacing

\bibitem{slime}
Y.~Chubachi and K.~Aiko, ``Slime: Automated anti-sandboxing disarmament
  system,''
  \url{https://www.blackhat.com/docs/asia-15/materials/asia-15-Chubachi-Slime-Automated-Anti-Sandboxing-Disarmament-System.pdf},
  2015.

\bibitem{deb2002fast}
K.~Deb, A.~Pratap, S.~Agarwal, and T.~Meyarivan, ``A fast and elitist
  multiobjective genetic algorithm: {NSGA-II},'' \emph{IEEE Transactions on
  Evolutionary Computation}, vol.~6, no.~2, pp. 182--197, 2002.

\bibitem{spider:acsac13}
Z.~Deng, X.~Zhang, and D.~Xu, ``Spider: Stealthy binary program instrumentation
  and debugging via hardware virtualization,'' in \emph{Proceedings of the
  Annual Computer Security Applications Conference (ACSAC'13)}, 2013.

\bibitem{deng2013spider}
------, ``Spider: stealthy binary program instrumentation and debugging via
  hardware virtualization,'' in \emph{Proceedings of the 29th Annual Computer
  Security Applications Conference}.\hskip 1em plus 0.5em minus 0.4em\relax
  ACM, 2013, pp. 289--298.

\bibitem{ether}
A.~Dinaburg, P.~Royal, M.~Sharif, and W.~Lee, ``{Ether: Malware analysis via
  hardware virtualization extensions},'' in \emph{Proceedings of the 15th ACM
  Conference on Computer and Communications Security (CCS '08)}, 2008.

\bibitem{sans-report}
D.~Distler, \emph{Malware Analysis: An Introduction}.\hskip 1em plus 0.5em
  minus 0.4em\relax SANS Institute, December 2007, available via
  \url{https://www.sans.org/reading-room/whitepapers/malicious/malware-analysis-introduction-2103}.

\bibitem{dolan2011virtuoso}
B.~Dolan-Gavitt, T.~Leek, M.~Zhivich, J.~Giffin, and W.~Lee, ``Virtuoso:
  Narrowing the semantic gap in virtual machine introspection,'' in
  \emph{Security and Privacy (SP), 2011 IEEE Symposium on}.\hskip 1em plus
  0.5em minus 0.4em\relax IEEE, 2011, pp. 297--312.

\bibitem{xen}
B.~Dragovic, K.~Fraser, S.~Hand, T.~Harris, A.~Ho, I.~Pratt, A.~Warfield,
  P.~Barham, and R.~Neugebauer, ``Xen and the art of virtualization,'' in
  \emph{In Proceedings of the ACM Symposium on Operating Systems Principles},
  2003.

\bibitem{anti-debug-symantec}
N.~Falliere, ``Windows anti-debug reference,''
  \url{http://www.symantec.com/connect/articles/windows-anti-debug-reference},
  2010.

\bibitem{forensic-discovery}
D.~Farmer and W.~Venema, \emph{Forensic Discover}.\hskip 1em plus 0.5em minus
  0.4em\relax Addison-Wesley, 2005.

\bibitem{hyperdbg:ase10}
A.~Fattori, R.~Paleari, L.~Martignoni, and M.~Monga, ``{Dynamic and Transparent
  Analysis of Commodity Production Systems},'' in \emph{Proceedings of the
  IEEE/ACM International Conference on Automated Software Engineering
  (ASE'10)}, 2010.

\bibitem{fratantonio16:triggerscope}
Y.~Fratantonio, A.~Bianchi, W.~Robertson, E.~Kirda, C.~Kruegel, and G.~Vigna,
  ``{TriggerScope: Towards Detecting Logic Bombs in Android Apps},'' in
  \emph{Proceedings of the IEEE Symposium on Security and Privacy (S\&P)}, San
  Jose, CA, May 2016.

\bibitem{hong2018}
Y.~Hong, Y.~Hu, C.-M. Lai, S.~Felix~Wu, I.~Neamtiu, P.~McDaniel, P.~Yu, H.~Cam,
  and G.-J. Ahn, ``Defining and detecting environment discrimination in android
  apps,'' in \emph{Security and Privacy in Communication Networks}, X.~Lin,
  A.~Ghorbani, K.~Ren, S.~Zhu, and A.~Zhang, Eds.\hskip 1em plus 0.5em minus
  0.4em\relax Cham: Springer International Publishing, 2018, pp. 510--529.

\bibitem{huijgens2017effort}
H.~Huijgens, A.~Van~Deursen, L.~L. Minku, and C.~Lokan, ``Effort and cost in
  software engineering: A comparison of two industrial data sets,'' in
  \emph{Proceedings of the 21st International Conference on Evaluation and
  Assessment in Software Engineering}.\hskip 1em plus 0.5em minus 0.4em\relax
  ACM, 2017, pp. 51--60.

\bibitem{idapro}
{IDA Pro}, \url{www.hex-rays.com/products/ida/}.

\bibitem{jain2014sok}
B.~Jain, M.~B. Baig, D.~Zhang, D.~E. Porter, and R.~Sion, ``Sok: Introspections
  on trust and the semantic gap,'' in \emph{2014 IEEE symposium on security and
  privacy}.\hskip 1em plus 0.5em minus 0.4em\relax IEEE, 2014, pp. 605--620.

\bibitem{jiang2007stealthy}
X.~Jiang, X.~Wang, and D.~Xu, ``Stealthy malware detection through vmm-based
  out-of-the-box semantic view reconstruction,'' in \emph{Proceedings of the
  14th ACM conference on Computer and communications security}.\hskip 1em plus
  0.5em minus 0.4em\relax ACM, 2007, pp. 128--138.

\bibitem{jiang}
\BIBentryALTinterwordspacing
------, ``Stealthy malware detection through vmm-based "out-of-the-box"
  semantic view reconstruction,'' in \emph{Proceedings of the 14th ACM
  Conference on Computer and Communications Security}, ser. CCS '07.\hskip 1em
  plus 0.5em minus 0.4em\relax New York, NY, USA: Association for Computing
  Machinery, 2007, p. 128–138. [Online]. Available:
  \url{https://doi.org/10.1145/1315245.1315262}
\BIBentrySTDinterwordspacing

\bibitem{Jing:2014:MAG:2664243.2664250}
\BIBentryALTinterwordspacing
Y.~Jing, Z.~Zhao, G.-J. Ahn, and H.~Hu, ``Morpheus: Automatically generating
  heuristics to detect android emulators,'' in \emph{Proceedings of the 30th
  Annual Computer Security Applications Conference}, ser. ACSAC '14.\hskip 1em
  plus 0.5em minus 0.4em\relax New York, NY, USA: ACM, 2014, pp. 216--225.
  [Online]. Available: \url{http://doi.acm.org/10.1145/2664243.2664250}
\BIBentrySTDinterwordspacing

\bibitem{JBremer2019}
{Jurriaan Bremer, Thorsten Sick, Rasmus Männa, and Mohsen Ahmadi},
  ``{VMCloak},'' \url{https://github.com/AdaptiveComputationLab/vmcloak}, 2020.

\bibitem{kaspersky-report-2017}
{Kaspersky Lab}, ``{Kaspersky Security Bulletin 2017},''
  \url{https://media.kaspersky.com/jp/pdf/pr/Kaspersky\_KSB2017\_Statistics-PR-1045.pdf}.

\bibitem{barebox-acsac11}
D.~Kirat, G.~Vigna, and C.~Kruegel, ``{BareBox: Efficient malware analysis on
  bare-metal},'' in \emph{Proceedings of the 27th Annual Computer Security
  Applications Conference (ACSAC'11)}, 2011.

\bibitem{kirat2014barecloud}
------, ``Barecloud: Bare-metal analysis-based evasive malware detection.'' in
  \emph{USENIX Security Symposium}, 2014, pp. 287--301.

\bibitem{kolbitsch2011power}
C.~Kolbitsch, E.~Kirda, and C.~Kruegel, ``The power of procrastination:
  detection and mitigation of execution-stalling malicious code,'' in
  \emph{Proceedings of the 18th ACM conference on Computer and communications
  security}.\hskip 1em plus 0.5em minus 0.4em\relax ACM, 2011, pp. 285--296.

\bibitem{draugr}
A.~Kopytov, ``Draugr---live memory forensics on linux,''
  \url{http://code.google.com/p/draugr}.

\bibitem{lengyel2014scalability}
T.~K. Lengyel, S.~Maresca, B.~D. Payne, G.~D. Webster, S.~Vogl, and A.~Kiayias,
  ``Scalability, fidelity and stealth in the drakvuf dynamic malware analysis
  system,'' in \emph{Proceedings of the 30th Annual Computer Security
  Applications Conference}, 2014, pp. 386--395.

\bibitem{qiang2018}
Q.~Li, Y.~Zhang, L.~Su, Y.~Wu, X.~Ma, and Z.~Yang, ``An improved method to
  unveil malware's hidden behavior,'' in \emph{Information Security and
  Cryptology}, X.~Chen, D.~Lin, and M.~Yung, Eds.\hskip 1em plus 0.5em minus
  0.4em\relax Cham: Springer International Publishing, 2018, pp. 362--382.

\bibitem{10.1007/978-3-642-23644-0_18}
M.~Lindorfer, C.~Kolbitsch, and P.~Milani~Comparetti, ``Detecting
  environment-sensitive malware,'' in \emph{Recent Advances in Intrusion
  Detection}, R.~Sommer, D.~Balzarotti, and G.~Maier, Eds.\hskip 1em plus 0.5em
  minus 0.4em\relax Berlin, Heidelberg: Springer Berlin Heidelberg, 2011, pp.
  338--357.

\bibitem{malwarebytes-report-2020}
{Malwarebytes Labs}, ``{State of Malware Report 2020},''
  \url{https://resources.malwarebytes.com/files/2020/02/2020_State-of-Malware-Report.pdf}.

\bibitem{mcafee4q-2014report}
{McAfee}, ``{Threats Report: Fourth Quarter 2014},''
  \url{http://www.mcafee.com/us/resources/reports/rp-quarterly-threat-q4-2014.pdf}.

\bibitem{mcafee1q-2016report}
------, ``{Threats Report: March 2016},''
  \url{http://www.mcafee.com/us/resources/reports/rp-quarterly-threats-mar-2016.pdf}.

\bibitem{miramir2017}
N.~Miramirkhani, M.~P. Appini, N.~Nikiforakis, and M.~Polychronakis, ``Spotless
  sandboxes: Evading malware analysis systems using wear-and-tear artifacts,''
  in \emph{2017 IEEE Symposium on Security and Privacy (SP)}, May 2017, pp.
  1009--1024.

\bibitem{miramirkhani2017spotless}
------, ``Spotless sandboxes: Evading malware analysis systems using
  wear-and-tear artifacts,'' in \emph{2017 IEEE Symposium on Security and
  Privacy (SP)}.\hskip 1em plus 0.5em minus 0.4em\relax IEEE, 2017, pp.
  1009--1024.

\bibitem{virtualbox}
{Oracle}, ``{VirtualBox},'' \url{http://www.virtualbox.com}, 2007.

\bibitem{pafish}
\BIBentryALTinterwordspacing
A.~Ortega, ``Paranoid fish.'' [Online]. Available:
  \url{http://github.com/a0rtega/pafish}
\BIBentrySTDinterwordspacing

\bibitem{oyama2018trends}
Y.~Oyama, ``Trends of anti-analysis operations of malwares observed in api call
  logs,'' \emph{Journal of Computer Virology and Hacking Techniques}, vol.~14,
  no.~1, pp. 69--85, 2018.

\bibitem{pan2017dark}
X.~Pan, X.~Wang, Y.~Duan, X.~Wang, and H.~Yin, ``Dark hazard: Learning-based,
  large-scale discovery of hidden sensitive operations in android apps.'' in
  \emph{NDSS}, 2017.

\bibitem{fatkit}
N.~L. Petroni, J.~Aaron, W.~Timothy, F.~William, and A.~Arbaugh, ``Fatkit: A
  framework for the extraction and analysis of digital forensic data from
  volatile system memory,'' \emph{Digital Investigation}, vol.~3, 2006.

\bibitem{anti-vm-ldt}
D.~Quist and V.~Val~Smith, ``{Detecting the Presence of Virtual Machines Using
  the Local Data Table},'' \url{http://www.offensivecomputing.net/}.

\bibitem{quist2006detecting}
D.~Quist, V.~Smith, and O.~Computing, ``Detecting the presence of virtual
  machines using the local data table,'' \emph{Offensive Computing}, 2006.

\bibitem{anti-emulator-ISC07}
T.~Raffetseder, C.~Kruegel, and E.~Kirda, ``Detecting system emulators,'' in
  \emph{Information Security}.\hskip 1em plus 0.5em minus 0.4em\relax Springer
  Berlin Heidelberg, 2007.

\bibitem{rlpack}
{Reversing Labs}, ``{RLPack},'' \url{https://reversinglabs.com}, retrieved
  November 2016.

\bibitem{redpill}
J.~Rutkowska, ``{Red Pill},'' \url{http://www.ouah.org/Red_Pill.html}.

\bibitem{saberi2014hybrid}
A.~Saberi, Y.~Fu, and Z.~Lin, ``Hybrid-bridge: Efficiently bridging the
  semantic gap in virtual machine introspection via decoupled execution and
  training memoization,'' in \emph{Proceedings of the 21st Annual Network and
  Distributed System Security Symposium (NDSS’14)}, 2014.

\bibitem{bitblaze}
D.~Song, D.~Brumley, H.~Yin, J.~Caballero, I.~Jager, M.~Kang, Z.~Liang,
  J.~Newsome, P.~Poosankam, and P.~Saxena, ``Bitblaze: A new approach to
  computer security via binary analysis,'' in \emph{Proceedings of the 4th
  International Conference on Information Systems Security (ICISS'08)}, 2008.

\bibitem{sonicwall-report-2020}
{Sonicwall}, ``{Sonicwall Cyber Threat Report 2020},''
  \url{https://www.sonicwall.com/resources/2020-cyber-threat-report-pdf/}.

\bibitem{lophi:ndss16}
C.~Spensky, H.~Hu, and K.~Leach., ``{LO-PHI: Low Observable Physical Host
  Instrumentation},'' in \emph{Proceedings of 2016 Network and Distributed
  System Security Symposium (NDSS'16)}, 2016.

\bibitem{lophi}
C.~Spensky, H.~Hu, and K.~Leach, ``{LO-PHI}: Low observable physical host
  instrumentation,'' in \emph{Networks and Distributed Systems Security
  Symposium 2016 (NDSS 2016)}, San Diego, CA, February 2016, acceptance rate:
  15.8\%.

\bibitem{sec-week-18}
S.~Stefnisson, ``Evasive malware now a commodity,''
  \url{https://www.securityweek.com/evasive-malware-now-commodity}, 2018.

\bibitem{symantec-istr2017}
{Symantec}, ``Internet security threat report,''
  \url{https://www.symantec.com/content/dam/symantec/docs/reports/istr-22-2017-en.pdf},
  2017.

\bibitem{symantec-report-2019}
{Symantec Labs}, ``{Internet Security Threat Report (ISTR) 2019},''
  \url{https://docs.broadcom.com/doc/istr-24-2019-en}, February 2019.

\bibitem{tanabe2018}
R.~Tanabe, W.~Ueno, K.~Ishii, K.~Yoshioka, T.~Matsumoto, T.~Kasama, D.~Inoue,
  and C.~Rossow, ``Evasive malware via identifier implanting,'' in
  \emph{Detection of Intrusions and Malware, and Vulnerability Assessment},
  C.~Giuffrida, S.~Bardin, and G.~Blanc, Eds.\hskip 1em plus 0.5em minus
  0.4em\relax Cham: Springer International Publishing, 2018, pp. 162--184.

\bibitem{tanabe2018evasive}
------, ``Evasive malware via identifier implanting,'' in \emph{International
  Conference on Detection of Intrusions and Malware, and Vulnerability
  Assessment}.\hskip 1em plus 0.5em minus 0.4em\relax Springer, 2018, pp.
  162--184.

\bibitem{vmware}
{VMWare, Inc.}, ``Vmware server,'' \url{http://www.vmware.com/products/server},
  2008.

\bibitem{langeland2008}
P.~L. Wedum, \emph{Malware Analysis: A Systematic Approach}.\hskip 1em plus
  0.5em minus 0.4em\relax Norwegian University of Science and Technology, 2008,
  master's Thesis: Available via
  {\url{https://brage.bibsys.no/xmlui//bitstream/handle/11250/261770/-1/347719_FULLTEXT01.pdf}}.

\bibitem{westphal2014vmi}
F.~Westphal, S.~Axelsson, C.~Neuhaus, and A.~Polze, ``Vmi-pl: A monitoring
  language for virtual platforms using virtual machine introspection,''
  \emph{Digital Investigation}, vol.~11, pp. S85--S94, 2014.

\bibitem{willems2012down}
C.~Willems, R.~Hund, A.~Fobian, D.~Felsch, T.~Holz, and A.~Vasudevan, ``Down to
  the bare metal: Using processor features for binary analysis,'' in
  \emph{Proceedings of the 28th Annual Computer Security Applications
  Conference}.\hskip 1em plus 0.5em minus 0.4em\relax ACM, 2012, pp. 189--198.

\bibitem{v2e-vee12}
\BIBentryALTinterwordspacing
L.-K. Yan, M.~Jayachandra, M.~Zhang, and H.~Yin, ``{V2E: Combining hardware
  virtualization and software emulation for transparent and extensible malware
  analysis},'' in \emph{Proceedings of the 8th ACM SIGPLAN/SIGOPS Conference on
  Virtual Execution Environments (VEE'12)}, 2012. [Online]. Available:
  \url{http://doi.acm.org/10.1145/2151024.2151053}
\BIBentrySTDinterwordspacing

\bibitem{ollydbg}
O.~Yuschuk, ``{OllyDbg},'' \url{www.ollydbg.de}.

\bibitem{malware-analysis1}
L.~Zelster, ``Mastering 4 stages of malware analysis,''
  \url{https://zeltser.com/mastering-4-stages-of-malware-analysis/}, February
  2015.

\bibitem{malt-sp15}
F.~Zhang, K.~Leach, H.~Wang, A.~Stavrou, and K.~Sun, ``{Using Hardware Features
  to Increase Debugging Transparency},'' in \emph{Proceedings of the 36th IEEE
  Symposium on Security and Privacy}, 2015.

\end{thebibliography}

\begin{IEEEbiography}[{\includegraphics[width=1in,height=1.25in,clip,keepaspectratio]{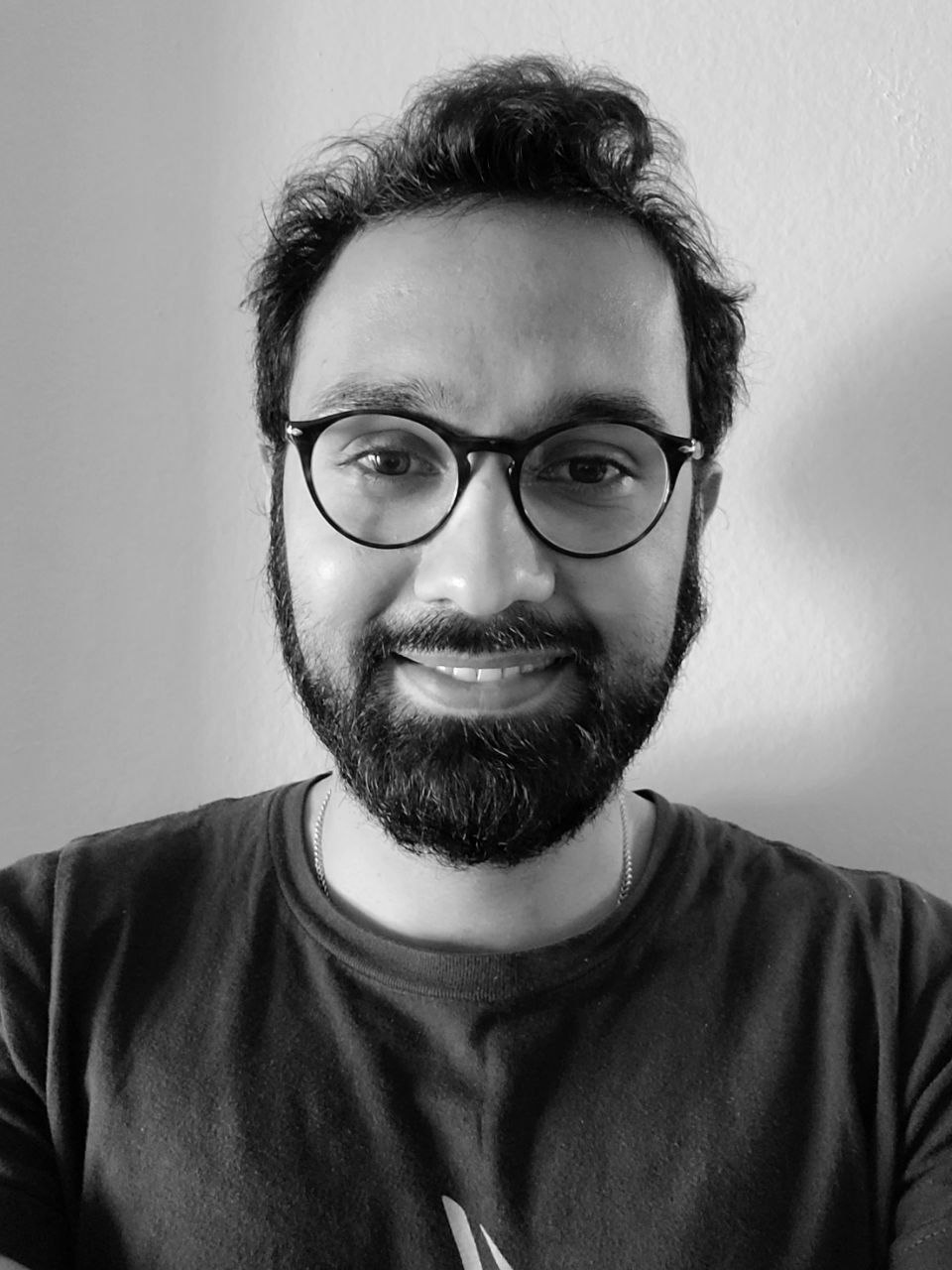}}]%
{Mohsen Ahmadi}
is a Senior Application Security Engineer. He received his MSc degree in Computer Science from Arizona State University and his BS from University of Isfahan. His main research is focused on program analysis, improving fuzzing techniques, and embedded security. He is a big open-source software fanatic and is an active security researcher.
\end{IEEEbiography}

\begin{IEEEbiography}[{\includegraphics[width=1in,height=1.25in,clip,keepaspectratio]{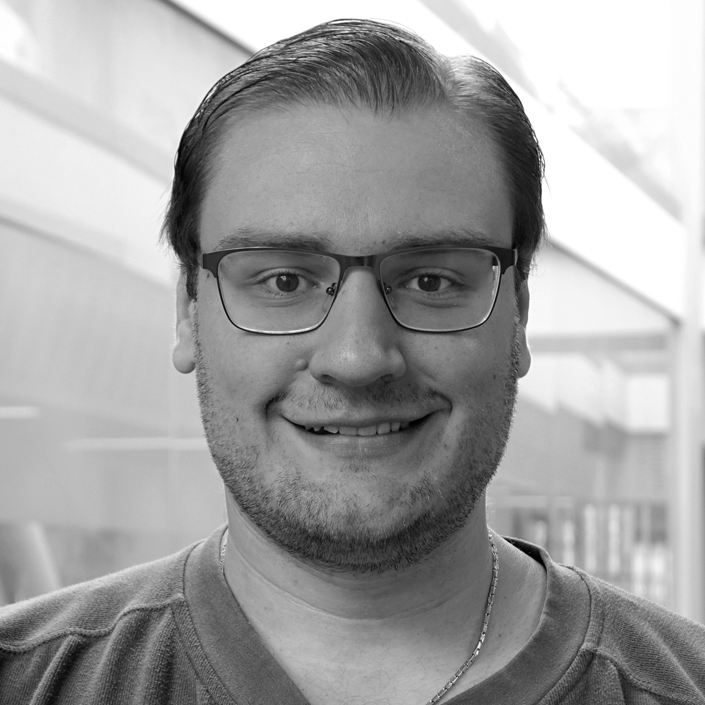}}]
{Kevin Leach} is a postdocotral researcher and lecturer at the University of Michigan.  
He earned the PhD from the University of Virginia in 2016.
His research combines the areas of systems security and software engineering --- he has developed techniques for transparent system introspection, kernel hotpatching, and stealthy malware analysis. 
\end{IEEEbiography}

\begin{IEEEbiography}[{\includegraphics[width=1in,height=1.25in,clip,keepaspectratio]{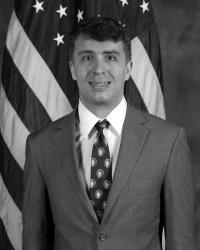}}]%
{Ryan Dougherty}
is an assistant professor in the Department of Electrical Engineering and
Computer Science at West Point. He earned his B.S. and Ph.D. from Arizona State University in 2019, and was a Visiting Assistant
Professor at Colgate University before joining West Point. His academic
interests include software engineering, theory of computation, and
combinatorial design theory.
\end{IEEEbiography}

\begin{IEEEbiography}[{\includegraphics[width=1in,height=1.25in,clip,keepaspectratio]{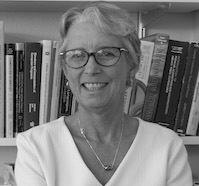}}]%
{Stephanie Forrest}
is a professor of Computer Science at Arizona State
University, where she directs the Biodesign Center for Biocomputation,
Security and Society.  Her interdisciplinary research focuses on the
intersection of biology and computation, including cybersecurity,
software engineering, and biological modeling.
\end{IEEEbiography}

\begin{IEEEbiography}[{\includegraphics[width=1in,height=1.25in,clip,keepaspectratio]{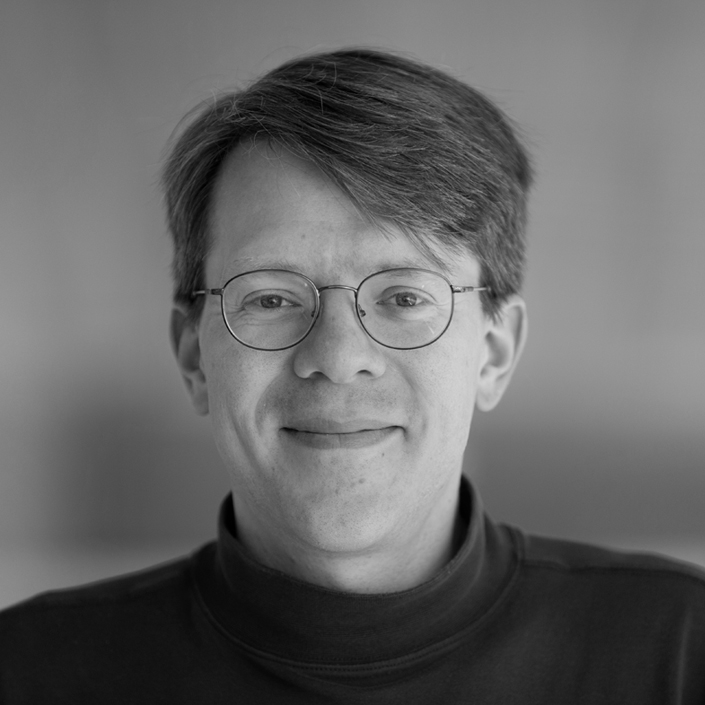}}]%
{Westley Weimer}
is a professor of Computer Science and Engineering at the University of Michigan. His main research interests include static, dynamic, and medical imaging-based techniques to improve program quality, fix defects, and understand how humans engineer software. He received a BA degree in computer science and mathematics from Cornell and MS and PhD degrees from Berkeley.
\end{IEEEbiography}

\end{document}